\documentclass[useAMS,usenatbib]{mn2e}
\usepackage{graphicx}
\usepackage{subfigure}
\usepackage[english]{babel}
\usepackage{amssymb}
\usepackage{xspace}
\def\hi{\ifmmode {\mbox H{\scshape i}}\else H{\scshape i}\fi\xspace}
\def\hii{\ifmmode {\mbox H{\scshape ii}}\else H{\scshape ii}\fi\xspace}
\def\h2{\ifmmode {\mbox H$_2$}\else H$_2$\fi\xspace}

\title [Evolution of atomic and molecular gas in galaxies]{Evolution
  of the atomic and molecular gas content of galaxies} 
\author[G. Popping, R.S. Somerville and
S.C. Trager]{Gerg\"o Popping$^{1}$\thanks{E-mail: g.popping@astro.rug.nl},
  Rachel S. Somerville$^{2}$ and Scott C. Trager$^{1}$ \\
$^{1}$Kapteyn Astronomical Institute, University of Groningen, Postbus 800, NL-9700 AV Groningen, the Netherlands\\
$^{2}$Department of Physics and Astronomy, Rutgers University, 136
Frelinghuysen Road, Piscataway, NJ 08854, USA}
\begin{document}

\maketitle

\begin{abstract}
We study the evolution of atomic and molecular gas in galaxies in
semi-analytic models of galaxy formation that include new modeling of
the partitioning of cold gas in galactic discs into atomic, molecular,
and ionised phases. We adopt two scenarios for the formation of
molecules: one pressure-based and one metallicity-based. We
find that both recipes successfully reproduce the gas fractions and
gas-to-stellar mass ratios of \hi and \h2 in local galaxies, as well
as the \hi and \h2 disc sizes up to $z \leq 2$. We reach good
agreement with the locally observed \hi and \h2 mass function,
although both recipes slightly overpredict the low-mass end of the \hi
mass function. Both of our models predict that the
high-mass end of the \hi mass function remains nearly constant at redshifts $z < 2.0$. The metallicity-based recipe yields a higher cosmic density of
cold gas and much lower cosmic \h2 fraction over the entire redshift
range probed than the pressure based recipe. These strong differences
in \hi mass function and cosmic density between the two recipes are
driven by low mass galaxies ($\log{(M_*/M_\odot)} \leq 7$) residing in
low mass halos ($\log{(M_{\rm{vir}}/M_\odot)} \leq 10$). Both recipes
predict that galaxy gas fractions remain high from $z \sim 6 -
3$ and drop rapidly at lower redshift. The galaxy \h2 fractions show a
similar trend, but drop even more rapidly. We provide predictions for the CO $J= 1-0$ luminosity of
  galaxies, which will be directly comparable with observations with
  sub-mm and radio instruments.
\end{abstract}


\section{Introduction}
Attaining an understanding of when, how, and at what rate stars form
out of interstellar gas, and of the mechanisms that regulate this
process, is of key importance in building up a complete picture of
galaxy formation and evolution. Observations across a range of scales
have shown that star-formation (SF) is tightly linked to galaxy gas
content. Observations in our Milky Way have shown that star formation
takes place in dense giant molecular clouds \citep[GMC;
  e.g.,][]{Solomon1987,McKee2007,Bolatto2008}. Early observational
work found a correlation between the surface density of the star
formation rate (SFR) and the surface density of the total cold gas in galaxies \citep[e.g.][]{Schmidt1959,Kennicutt1998KS}, while
more recent work has emphasized that there is a strong correlation
between the SFR density and the density of \emph{molecular} hydrogen
(\h2), while the correlation with atomic hydrogen (\hi) is weak or
absent \citep{wong2002,Bigiel2008,Bigiel2011,Schruba2011}.  This work
has stimulated a desire to understand and track the \hi\ and
\h2\ content of galaxies separately in theoretical models.

Observational studies of the \hi\ and \h2 content of nearby galaxies
have made great advances in recent years. The local \hi\ mass function
down to masses of $\log{(M_{\rm{HI}}/M_\odot)} = 7$, and global \hi
density $\Omega_{\rm HI}$, has been quantified by blind surveys such
as ALFALFA \citep{Giovanelli2005, Martin2010}. The \hi
content and its relationship with other galaxy properties (such as
stellar mass, stellar surface density, color, and concentration) have
been quantified for a fairly large, homogeneously selected sample of
nearby galaxies by GASS \citep[Galex Arecibo SDSS
  survey:][]{Catinella2010,Catinella2012,Catinella2013}.
The THINGS \citep[The \hi nearby galaxy survey:][]{Walter2008} and
LITTLE THINGS \citep{Hunter2012} surveys mapped the atomic
hydrogen density distribution in small samples of nearby galaxies.

The molecular hydrogen content of galaxies has most commonly been
studied through emission from $^{12}$CO (from here on CO) as a
tracer. The CO mass-function of nearby galaxies was presented by
\citet{Keres2003}, along with an estimate of the \h2 mass-function
resulting from the application of a constant conversion factor between
CO luminosity and \h2 mass. An updated estimate of the \h2 mass
function from the \citet{Keres2003} sample, based on an empirical, and
variable, CO-\h2 conversion factor, was presented by
\citet{Obreschkow2009}. The BIMA SONG \citep[BIMA survey of nearby
  galaxies:][]{Helfer2003}, HERACLES \citep[HERA CO-Line Extragalactic
  Survey:][]{Leroy2009_HERACLES} and COLD GASS \citep[CO legacy
  database for GASS:][]{Saintonge2011} surveys mapped the CO-emitting
gas in galaxies of the THINGS and GASS surveys, constraining the
surface densities and gas-to-star ratios of molecular gas.

Observations of atomic hydrogen in emission have up until now been
restricted to galaxies at redshifts of $z\lesssim 0.2$
\citep{Verheijen2007,Catinella2008}. Damped Lyman-$\alpha$ absorbers
(DLA) have provided estimates of the global atomic gas content of the
Universe ($\Omega_{\rm gas}$) at much higher redshifts \citep[$z <
  4.5$; e.g., ][]{Rao2006,Prochaska2009,Noterdaeme2012}, but the exact
nature of these systems, and their connection to galaxies detected in
emission, is still unclear, making the interpretation of these
observations somewhat complicated \citep{Berry2013}.

Direct observations of the molecular content of distant galaxies
through the CO line have recently become available for small samples
of objects, although these samples are usually biased towards the most
gas-rich, actively star-forming galaxies \citep[e.g.][]{Genzel2010, Tacconi2010,
Riechers2011, Bothwell2013,Tacconi2013}.  Although results are still inconclusive
because of the small and potentially biased nature of the samples, and
uncertainties in the \h2-CO conversion factor, these studies suggest
that galaxies at high redshift may have been considerably more rich in
\h2 than nearby galaxies. Moreover, a tight relationship between \h2
surface density and SFR density seems to persist out to at least
$z\sim 2$ \citep{Genzel2010, Daddi2010}.

The gas content of galaxies at high redshift has also been estimated
using more indirect methods, such as by using Far-Infrared (FIR)
observations and an assumed relationship between dust and \h2 mass
\citep{Magdis2012}, or by using an empirical relationship between SFR
density and total gas or \h2 content along with SF tracers such as
H-$\alpha$ or UV \citep{Erb2006,Mannucci2009,Popping2012}. 

All of the above efforts have already led to extremely valuable
insights and constraints on galaxy formation models. However, our
ability to measure \hi and CO in emission, in unbiased samples of
galaxies out to high redshift, is expected to undergo a revolution in
the next decade, with new and upcoming facilities such as the Atacama Large Millimeter Array
\citep[ALMA;][]{Wootten2009} and the Square Kilometer Array \citep[SKA;][]{Carilli2004}
and its pathfinders the Karoo Array Telescope
\citep[MeerKAT;][]{Booth2009} and the Australian SKA Pathfinder
\citep[ASKAP;][]{Johnston2008} coming online.

The observations expected from these facilities present a new and
stringent challenge to theoretical models of galaxy formation. Until
recently, most cosmological models and simulations of galaxy formation
did not attempt to `partition' gas into different phases, and used a
total-gas based (\hi+\h2) Kennicutt-Schmidt (KS) law to model star
formation. However, aided by the insights gained from studies of the
relationship between star formation and gas properties on $\sim$ kpc
scales \citep[e.g.][]{Bigiel2008,Leroy2008} in external galaxies, theorists have also made
considerable progress on developing physical models linking the
efficiency of star formation on GMC scales with that on galactic
scales. Several groups have implemented explicit modeling of detailed
chemistry and simplified radiative transfer into galaxy-scale and
cosmological numerical hydrodynamic simulations, tracking the
multi-phase gas content and implementing \h2-based star formation
prescriptions \citep[e.g.][]{Pelupessy2006,Robertson2008,Gnedin2011,Christensen2012,Kuhlen2012}.
\citet[][hereafter GK]{Gnedin2011} presented fitting functions for the
SFR in their simulations as a function of total cold gas density
($\Sigma_{\rm HI}+\Sigma_{\rm H_2}$), gas phase metallicity, and the
intensity of the UV ionizing background. \citet{Krumholz2009} presented
analytic models for the formation of \h2 as a function of total gas
density and metallicity, supported by numerical simulations with
simplified geometries \citep{Krumholz2008,Krumholz2009}, emphasizing the
importance of metallicity as a controlling parameter in
\h2\ formation. A somewhat different view is presented by
\citet{Ostriker2010}, who propose that heating of the Interstellar
Medium (ISM) by the stellar UV background plays a key role in
regulating star formation. In their model, the thermal pressure in the
diffuse ISM, which is proportional to the UV heating rate, adjusts
until it balances the midplane pressure set by the vertical
gravitational potential. This could provide an explanation for the
strong empirical correlation between \h2 fraction and disc midplane
pressure found by \citet{Blitz2006}.

These analytic models and fitting formulae can be implemented within
semi-analytic models of galaxy formation. The modern semi-analytic
approach applies simple, physically motivated recipes for physical
processes that drive the formation and evolution of galaxies within
the framework of a $\Lambda$ cold dark matter ($\Lambda$CDM) cosmology. These models
can provide predictions of global galaxy properties (such as SFR,
size, stellar mass and luminosity, gas content, metal enrichment) for
large numbers of galaxies. Furthermore, they can efficiently explore
the parameter space associated with the large number of ``sub-grid''
recipes that are used to model processes such as star formation,
stellar feedback, black hole accretion and feedback from Active
Galactic Nuclei (AGN). Semi-analytic models have been successful in
reproducing many observed galaxy properties
\citep[e.g.,][]{Kauffmann1993,Cole1994,Kauffmann1999,Somerville1999,Cole2000,Somerville2001},
in particular recent models that include `radio mode' AGN feedback
\citep[e.g.,][]{Bower2006,Croton2006,Kang2006,Menci2006,Monaco2007,Somerville2008},
although some puzzles remain. For example, SAMs from several different
groups do not correctly reproduce the observed properties of low-mass
galaxies \citep[$\log{(M_*/M_\odot)} \sim 9 - 10.5$][]{Fontanot2009,Guo2010,Weinmann2012}. These low-mass galaxies
form too early in the models, and are too passive at late times. On
the other hand, SAMs have also had difficulty reproducing enough very
rapidly star forming, extreme starbursts (Ultra-luminous Infrared
Galaxies; ULIRGS) at high redshift \citep[and references therein]{Somerville2012,Niemi2012}. However, numerical hydrodynamic
simulations suffer from the same problems
\citep{Weinmann2012,Dave2010}, and in fact produce very similar
predictions to the SAMs, leading most theorists to conclude that it is
likely to be limitations in our understanding of the sub-grid
processes of star formation and stellar feedback, rather than
inaccuracies of the semi-analytic approach, that are the root cause of
the problems.

Several groups have now used semi-analytic models to make predictions
for the multi-phase gas content of
galaxies. \citet{Obreschkow2009_sam} applied an empirical
pressure-based recipe based on the results of \citet[][hereafter
  BR]{Blitz2006} in post-processing to compute the \hi and \h2 content
of galaxies in the Millennium simulations
\citep{DeLucia2007}. \citet{Power2010} carried out a similar project
based on post-processing. \citet{Fu2010,Fu2012} modeled the partitioning of gas into
\hi and \h2 in radial bins in each galaxy, using both the
metallicity-dependent recipes of \citet[][hereafter KMT]{Krumholz2009} and
the pressure-based recipe of BR, and self-consistently implemented a
\h2-based star formation recipe, within the established semi-analytic
modeling framework of
\citet{Guo2011}. \citet{Lagos2011cosmic_evol,Lagos2011sflaw} also
estimated gas partitioning into an atomic and molecular component, and
implemented a \h2-based star formation recipe, within the GALFORM
semi-analytic model \citep{Baugh2005,Bower2006}. Somewhat simpler models
in a similar spirit have also been presented by \citet{Dutton2010}
and \citet{Krumholz2011}. 

In this paper we explore how different models for \h2 formation
  affect the evolution of the atomic and molecular gas content of
  galaxies with time. We implement an empirical, pressure-based recipe
  (BR) and a recipe based on numerical hydrodynamic simulations,
  dependent on metallicity and the local UV radiation field (GK) into the
  \citet{Somerville2012} model, thus allowing a link to be made between the stellar and dust
emission and the multi-phase gas content of galaxies. We anticipate
that these predictions will be useful for planning upcoming
observations of cold gas in galaxies at high redshift, and as these
observations become available, this will provide insights into the
physics that drives the formation of molecules in
galaxies. Furthermore, we aim to give insight to what improvements
need to be incorporated in cosmological galaxy evolution models to
correctly model the gas content of galaxies. In Somerville, Popping \&
Trager (in prep.; SPT14) we implement a wider suit of star-formation
and \h2 recipes including the KMT recipe. We will present predictions for the observable
 properties of the \emph{stellar} (and dust) emission over a broad
 range of redshifts, and discuss the sensitivity of these properties
 to the adopted SF-recipes.

The structure of the paper is as follows. In Section \ref{sec:model}
we briefly present the semi-analytic model and its ingredients,
focussing on the new recipes for gas partitioning and star formation. In
Section \ref{sec:results} we present our predictions for the scaling
relations between stellar mass or surface density and \hi and \h2
content, relationship between \hi mass and radius, and \hi and \h2
mass functions at $z\sim 0$. We further present predictions for the
evolution in the SFR half-light radius vs. stellar mass, \hi and \h2
mass functions, global mass density of \hi and \h2, and \hi and \h2
fractions vs. stellar mass. We compare our predictions of \h2
fractions and mass functions with observational estimates of these
quantities obtained by applying a CO-\h2 conversion factor to the
observations; we also adopt an alternate approach in which we use our
knowledge of the physical properties of our model galaxies to estimate
the CO content, and compare directly with the CO observations. In Section
\ref{sec:discussion} we discuss our findings and we summarize those in
Section \ref{sec:conclusion}.

\section{Model}
\label{sec:model}

This section describes the semi-analytic model used in this paper. The
model is based on the models presented in \citet{Somerville1999},
\citet{Somerville2008}, and \citet{Somerville2012} and we refer the reader to those papers for
details. In this section we provide a brief summary of the model
framework and the ingredients relevant to this paper.  Throughout this
paper we adopt a flat $\Lambda$CDM cosmology with $\Omega_0 =
0.28,\,\Omega_\Lambda = 0.72, h = H_0/(100\,
\rm{km}\,\rm{s}^{-1})=0.70, \sigma_8 = 0.812$ and a cosmic baryon
fraction of $f_b = 0.1658$ \citep{Komatsu2009}. Unless stated otherwise we leave the free parameters associated
with the galaxy-formation model fixed to the values given in
\citet{Somerville2012}.

\subsection{Semi-analytic model framework}
The merging histories of dark matter halos (merger trees) are
constructed based on the Extended Press-Schechter formalism following
the method described in \citet{Somerville99tree} and
\citet{Somerville2008}. Each branch in the tree represents a merger
event and is followed back in time to a minimum progenitor mass of
$M_{\rm{res}}$, which we refer to as the mass resolution of our
simulations.

Whenever dark matter halos merge, the central galaxy of the largest
progenitor halo becomes the new central galaxy, whereas all the other
galaxies become `satellites'. Satellite galaxies may eventually
merge with the central galaxy due to dynamical friction. Merger
timescale are estimated using a variant of the Chandrasekhar formula from
\citet{Boylan2008}. Tidal stripping and destruction of the satellites
is included as described in \citet{Somerville2008}.

Before reionisation of the Universe, each halo contains a mass of hot
gas equal to the universal baryon fraction times the virial mass of
the halo. After reionisation, the collapse of gas into low-mass halos
is suppressed by the photoionising background. We model the fraction
of baryons that can collapse into halos of a given mass after
reionisation using the fitting functions provided by
\citet{Gnedin2000} and \citet{Kravtsov2004}.

When a dark matter halo collapses or experiences a merger with a
larger halo, the hot gas shock-heats to the virial temperature of the
new halo. The radiating gas then gradually cools and collapses. To
calculate the cooling rate of the hot gas we use the
metallicity-dependent radiative cooling curves of
\citet{sutherland1993}. The rate at which gas can cool is given by
\begin{equation}
\dot{m}_{\mathrm{cool}}=\frac{1}{2}m_{\mathrm{hot}}\frac{r_{\mathrm{cool}}}{r_{\mathrm{vir}}}\frac{1}{t_{\mathrm{cool}}},
\end{equation}
where $m_{\mathrm{hot}}$ is the mass of the hot halo gas,
$r_{\mathrm{vir}}$ is the virial radius of the dark matter halo, and
$r_{\mathrm{cool}}$ is the radius within which all of the gas can cool
in a time $t_{\mathrm{cool}}$, which itself depends on density,
metallicity and temperature. This cooling radius limited regime is
associated with ``hot flows''. In some cases the cooling radius can be
larger than the virial radius. In this case the cooling rate is
limited by the infall rate
\begin{equation}
\dot{m}_{\mathrm{cool}}=\frac{1}{2}m_{\mathrm{hot}}\frac{1}{t_{\mathrm{cool}}}.
\end{equation}
This infall limited cooling regime is associated with ``cold flows''
\citep{Birnboim2003,Keres2005,Dekel2006}.

Although in reality satellite galaxies should continue to accrete some
cold gas, we assume that the cold gas is only accreted by the central
galaxy of the halo. When the gas cools we assume it initially
collapses to form a rotationally supported disc. The scale radius of
the disc is computed based on the initial angular momentum of the gas
and the halo profile, assuming that angular momentum is conserved and
that the self-gravity of the collapsing baryons causes contraction of
the matter in the inner part of the halo
\citep{Blumenthal1986,Flores1993,Mo1998}. Assuming that the
  halo initially has a density profile described by the
  Navarro-Frank-White \citep[NFW;][]{Navarro1996} form, the size of
  the gas disc of a galaxy is given by
\begin{equation}
r_{\rm gas} = \frac{1}{\sqrt{2}}f_j\lambda R_{\rm vir}f_c^{-1/2}f_{\rm R}(\lambda,c,f_d),
\end{equation}
where $f_j \equiv (J_d/m_d)/(J_h/M_{\rm{vir}})$ is the ratio of the
specific angular momentum of the disc and the halo, $c$ is the NFW
concentration of the halo, and $f_d$ is the disc mass to the halo mass
ratio. The functions $f_c^{-1/2}$ correct for the difference in energy
of the NFW profile relative to that of a singular isothermal profile,
and $f_{\rm R}$ accounts for the adiabatic contraction \citep[see][for
  expressions governing $f_{\rm R}$ and $f_c$]{Mo1998}.
\citet{Somerville2008size} showed that this approach produced good
agreement with the evolution of the size-stellar mass relation for
disc-dominated galaxies from $z\sim 2$ to the present.

Stars are formed through two modes, a ``normal'' mode in isolated
discs, and a merger-driven ``starburst'' mode. We discuss star
formation in the ``normal'' mode in below Section \ref{sec:h2_sf}. The
efficiency and timescale of the ``starburst'' mode is set by the merger
mass ratio and the gas fractions of the merger progenitors, based on
the results of hydrodynamical simulations of binary galaxies
\citep{Robertson2006,Hopkins2009a}.

When supernovae occur, they deposit some of their energy into the ISM, driving
a large-scale outflow of cold gas from the galaxy. The mass outflow
rate is given by
\begin{equation}
\dot{m}_{\rm out} = \epsilon_{\rm SN} \left(\frac{V_0}{V_c} \right)^{\alpha_{\rm rh}} \dot{m}_*
\end{equation}
where $V_c$ is the maximum circular velocity of the galaxy (here
approximated by $V_{\rm max}$ of the uncontracted dark matter halo),
$\dot{m}_*$ is the star formation rate, and $\epsilon_{\rm SN}$ and
$\alpha_{\rm SN}$ are free parameters ($V_0=200$ km/s is an arbitrary
normalization constant). Some fraction of the ejected gas escapes from
the potential of the dark matter halo, whereas some is deposited in
the hot gas reservoir within the halo and can cool again. The fraction
of gas ejected from the disc and halo versus ejected from the disc but
retained in the halo is a function of the halo circular velocity, such
that low-mass halos lose a larger fraction of gas (see
\citet{Somerville2008} for details).  We choose $\epsilon_{\rm
    SN} = 1.5$ and $\alpha_{\rm SN} = 2.2$ (similar to previous works)
  to obtain a good match with the observed $z\sim 0.0$ stellar mass
  function.

Each generation of stars produces heavy elements that can enhance the
metal content of a galaxy. Here, chemical enrichment is modelled in a
simple manner using the instantaneous recycling approximation. For
each parcel of new stars ${\rm d}m_*$, we also create a mass of metals
${\rm d}M_Z = y \, {\rm d}m_*$, which we assume to be instantaneously
mixed with the cold gas in the disc. We assume the yield $y$ to be
constant, and treat it as a free parameter. When supernova driven
winds eject gas from the disc, a corresponding proportion of metals is
also removed and deposited either in the hot gas or outside the halo,
following the same proportions as the ejected gas.

Mergers can remove angular momentum from the disc stars and build up a
spheroid. The efficiency of disc destruction and build up of spheroids
is a function of progenitor merger mass ratio and gas fractions,
parameterised based on the simulations of binary galaxy systems
\citep{Hopkins2009a}. These simulations indicate that more ``major''
and more gas-poor mergers are more efficient in removing angular
momentum, destroying discs, and building spheroids. When implemented
within the SAM, these recipes correctly predict the relative fractions
of early vs. late type galaxies as a function of stellar mass
\citep{Hopkins2009b}. 

The model tracks the growth of supermassive black holes and the energy
they release \citep{Croton2006,Somerville2008}. Each top-level DM halo is seeded with a $\sim 100
M_\odot$ black hole, and these black holes are able to grow via two
different accretion modes. The first accretion mode is fueled by cold
gas that is driven into the nucleus of the galaxy by mergers. This
mode is radiatively efficient, and the accretion rates are close to
the Eddington limit. The accretion continues until the energy being
deposited into the ISM in the central region of the galaxy is
sufficient to significantly offset and halt accretion via a
pressure-drive outflow. Because this accretion mode is associated with
optically bright classical quasars and AGN, it is sometimes referred to as
``bright mode'' or ``quasar mode'' accretion. The second mode of black hole growth, the ``radio mode'', is thought
to be associated with powerful jets observed at radio frequencies. Hot
halo gas is assumed to be accreted according to the Bondi-Hoyle
approximation \citep{Bondi1952}. We adjust the efficiency of
  ``radio mode'' heating to fit the observed number density of massive
  galaxies, and obtain $\kappa_{\rm radio} = 3.8 \times 10^{-3}$).
Accretion rates in this mode are significantly sub-Eddington
($10^{-4}$ to $10^{-3}$ times the Eddington rate), so that most of the
BH's mass is acquired during ``bright mode'' accretion. However, the
radio jets are assumed to couple very efficiently with the hot gas,
and provide a heating term that can partially or completely offset
cooling during the ``hot flow'' mode.

\subsection{Multi-phase Gas Partitioning and Star Formation Recipes}
\label{sec:h2_sf}
In this section we describe the new ingredients of our model that we
use to calculate the fraction of ionised, atomic, and molecular gas in
each galaxy, and how we compute the SFR based on the molecular gas
content. 

At each time step we compute the scale radius of the cold gas disc
using the angular momentum argument described in the previous
subsection. We assume that the cold gas is distributed in an
exponential disc with scale radius $r_{\mathrm{gas}}$ and a central gas surface density of $m_{\rm
    cold}/(2\pi\,r_{\mathrm{gas}}^2)$, where $m_{\rm cold}$ is the
mass of all cold gas in the disc. We assume that the stellar scale
length is defined as $r_{\mathrm{star}} =
r_{\mathrm{gas}}/\chi_{\mathrm{gas}}$, with $\chi_{\mathrm{gas}}=1.7$
fixed to match stellar scale lengths at $z = 0$. We divide the gas
disc into radial annuli and compute the fraction of molecular gas in
each annulus as described below. The integrated mass of \hi and \h2 in
the disc at each time step is calculated using a fifth order
Runga-Kutta integration scheme.

\subsubsection{Ionised gas}
We assume that the cold gas consists of an ionised, atomic and
molecular component. The ionised component may be due to either an
external background or by the radiation field from stars within the
galaxy. We assume that some fraction of the cold gas in the galaxy,
$f_{\rm ion, int}$, is ionised by the stars in the galaxy. The
external background field ionises of a slab of gas on each side of the
disc. Following \citet{Gnedin2012}, and assuming that all the
gas with a surface density below some critical value $\Sigma_{\rm
  HII}$ is ionised, we  use
\begin{equation}
 f_{\rm ion} = \frac{\Sigma_{\rm HII}}{\Sigma_0}
\left[1 + \ln \left(\frac{\Sigma_0}{\Sigma_{\rm HII}} \right) + 0.5
  \left(\ln \left(\frac{\Sigma_0}{\Sigma_{\rm HII}}\right) \right)^2
  \right]. 
\end{equation}
Throughout this paper we assume $f_{\rm ion, int} = 0.2$ (as in the
Milky Way) and $\Sigma_{\rm HII} = 0.4 \, M_\odot \rm{pc}^{-2}$,
supported by the results of \citet{Gnedin2012}. Although
  observations do not support a sharp transition to ionized gas at
  this surface density, we found that our model reproduced the results
  of the hydro simulations well with this choice of parameters.

\subsubsection{Molecular gas: pressure based partitioning}
In this work we consider two approaches for calculating the molecular
fraction of the cold neutral gas in a galaxy. The first is based on
the empirical pressure-based recipe presented by
\citet[BR]{Blitz2006}.  They found a power-law relation between the
disc mid-plane pressure and the ratio between molecular and atomic
hydrogen, i.e.,
\begin{equation}
R_{\mathrm{H}_2} = \bigl(\frac{\Sigma_{\mathrm{H}_2}}{\Sigma_{\mathrm{HI}}}\bigr) = \bigl(\frac{P_m}{P_0}\bigr)^\alpha
\label{eq:blitz2006}
\end{equation}
where $\Sigma_{\mathrm{H}_2}$ and $\Sigma_{\mathrm{HI}}$ are the \h2
and \hi surface density, $P_0$ and $\alpha_{\rm BR}$ are free parameters that are
obtained from a fit to the observational data, and $P_m$ the mid-plane
pressure acting on the galactic disc. 
We adopted $\log P_0/k_B = 4.23$ cm$^3$ K and $\alpha_{\rm BR}=0.8$
from \citet{Leroy2008}.  The hydrostatic pressure acting on the disc
at a radius $r$ is estimated as
\citep{Elmegreen1989,Elmegreen1993,Fu2010}
\begin{equation}
P_m(r) = \frac{\pi}{2}\,G\,\Sigma_{\mathrm{gas}}(r)\left[\Sigma_{\mathrm{gas}}(r) + f_{\sigma}(r)\Sigma_*(r)\right]
\label{eq:pressure}
\end{equation}
where G is the gravitational constant, $f_\sigma(r)$ is the
  ratio between $\sigma_{\mathrm{gas}}(r)$ and $\sigma_*(r)$, the gas
  and stellar vertical velocity dispersion, respectively. The stellar
  surface density profile $\Sigma_*(r)$ is modeled as an exponential
  with scale radius $r_{\mathrm{star}}$ and central density
  $\Sigma_{*, 0} \equiv m_*/(2 \pi r_*^2)$.  Following \citet{Fu2010},
  we adopt $f_{\sigma}(r) = 0.1 \sqrt{\Sigma_{*,0}/\Sigma_*}$, based
  on empirical scalings for nearby disc galaxies. The fraction of
  non-ionized gas in a molecular state at each radial annulus can be
  calculated as $f_{\rm{H}_2} = R_{\rm{H}_2}/(1 + R_{\rm{H}_2})$.

\subsubsection{Molecular gas: metallicity based partitioning}
The second approach for computing molecular gas fractions in galaxies
is based on the simulation by \citet[GK]{Gnedin2011}, who performed
high-resolution ``zoom-in'' cosmological simulations with the Adaptive
Refinement Tree (ART) code of \citet{Kravtsov99}, including gravity,
hydrodynamics, non-equilibrium chemistry, and 3D on the fly radiative
transfer. Based on their simulations, the authors find a fitting
function for the \h2 fraction which effectively parameterizes $f_{\rm
  H2}$ as a function of dust-to-gas ratio relative to the Milky Way,
$D_{\rm MW}$, the UV ionizing background relative to the Milky Way,
$U_{\rm MW}$, and the neutral gas surface density
$\Sigma_{HI+H_2}$. The fraction of molecular hydrogen at each radial
annulus is given by
\begin{equation}
 f_{H_2}(r) = \left[1+\frac{\tilde{\Sigma}}{\Sigma_{HI+H_2}(r)}\right]^{-2} 
\end{equation}
where
\begin{eqnarray*}
\tilde{\Sigma} & = &  20\, {\rm M_\odot pc^{-2}} \frac{\Lambda^{4/7}}{D_{\rm MW}} 
\frac{1}{\sqrt{1+U_{\rm MW} D_{\rm MW}^2}} \\
\Lambda & = & \ln(1+g D_{\rm MW}^{3/7}(U_{\rm MW}/15)^{4/7})\\
g & = & \frac{1+\alpha s + s^2}{1+s}\\
s &  = & \frac{0.04}{D_*+D_{\rm MW}}\\
\alpha &  = & 5 \frac{U_{\rm MW}/2}{1+(U_{\rm MW}/2)^2}\\
D_* & = & 1.5 \times 10^{-3} \, \ln(1+(3U_{\rm MW})^{1.7})
\end{eqnarray*}
We take the dust-to-gas ratio to be proportional to the metallicity in
solar units $D_{\rm MW} = Z/Z_{\odot}$. The local UV
  background relative to the MW is set by relating the SFR of the
  galaxy in the previous time step to the MW SFR as $U_{\rm MW}=
  \frac{SFR}{SFR_{\rm MW}}$, where we choose $SFR_{\rm MW} =
  1.0\,\rm{M}_\odot\,\rm{yr}^{-1}$ \citep{Murray2010,Robitaille2010}.

The GK fitting functions are intended to characterize the
  formation of molecular hydrogen on dust grains, the dominant
  mechanism for forming \h2 once gas is enriched to more than a few
  hundredths of Solar metallicity. Other channels for the formation of
  \h2 in primordial gas must be responsible for producing the
  molecular hydrogen out of which the first stars were
  formed. Hydrodynamic simulations containing detailed chemical
  networks and analytic calculations have shown that \h2 can form
  through other channels in dark matter halos above a critical mass
  $M_{\rm crit} \sim 10^5 \rm{M}_\odot$
  \citep[e.g.,][]{Nakamura2001,Glover2013}. This gas can then form
  ``Pop III'' stars which can enrich the surrounding ISM to
  $\rm{Z}_{\rm III} \sim 10^{-3}\,\rm{Z}_\odot$
  \citep{Schneider2002,Greif2010,Wise2012}. These processes take place
  in halos much smaller than our resolution limit. We represent them
  by setting a ``floor'' to the molecular hydrogen fraction in our
  halos, $f_{\rm H2,floor}$. In addition, we ``pre-enrich'' the
  initial hot gas in halos, and the gas accreted onto halos due to
  cosmological infall, to a metallicity of $\rm{Z}_{\rm
    pre-enrich}$. We adopt typical values of $f_{\rm H2,floor} =
  10^{-4}$ and $\rm{Z}_{\rm pre-enrich}=10^{-3}\rm{Z}_\odot$
  \citep{Haiman1996,Bromm2004}. We find that our results are not
  sensitive to the adopted values of these parameters within
  reasonable limits. Note that observations of resolved stars in the
  MW halo and local dwarfs have revealed stars with metallicities
  below $\rm{Z}\sim10^{-3}\, \rm{Z}_\odot$
  \citep{Tolstoy2009,Starkenburg2010}, precluding much higher values
  for $\rm{Z}_{\rm pre-enrich}$. 

\begin{figure}
\includegraphics[width = 1.0\hsize]{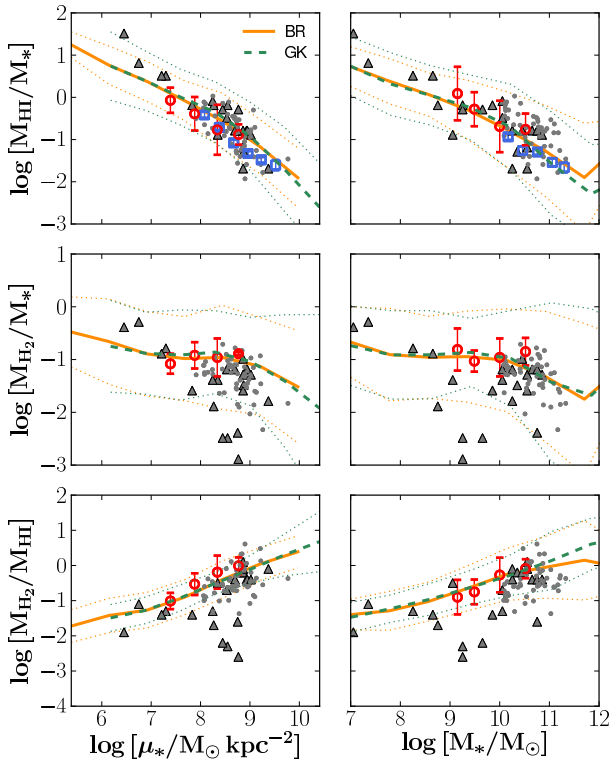}
\caption{Scaling relations for the \hi and \h2 content of
  disc-dominated galaxies ($M_{*_{\rm{bulge}}}/M_{*_{\rm{total}}} \leq
  0.4$) as a function of stellar mass and stellar surface density for
  the pressure- (solid orange) and metallicity-based (green dashed)
  \h2 formation recipes. Thick lines show the mean values and
    dotted lines mark the $2\sigma$ deviation. Grey triangles and
  dots, blue squares, and red circles are literature values from
  \citet{Leroy2008}, \citet{Saintonge2011}, \citet{Catinella2013}, and
  \citet{Boselli2014}, respectively.\label{fig:gas_scaling}}
\end{figure}
\begin{figure}
\includegraphics[width = 1.0\hsize]{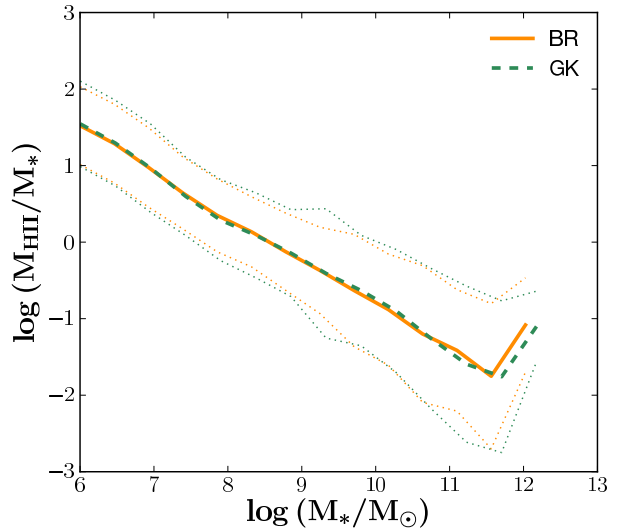}
\caption{The ratio of ionised hydrogen to stellar mass in
  disc-dominated galaxies ($M_{*_{\rm{bulge}}}/M_{*_{\rm{total}}} \leq
  0.4$) as a function of stellar mass for the pressure- (solid orange)
  and metallicity-based (dashed green) \h2 formation recipes. Thick lines show the mean, and dotted lines mark the $2\sigma$
    deviation.\label{fig:HII}}
\end{figure}
\begin{figure}
\includegraphics[width = 1.0\hsize]{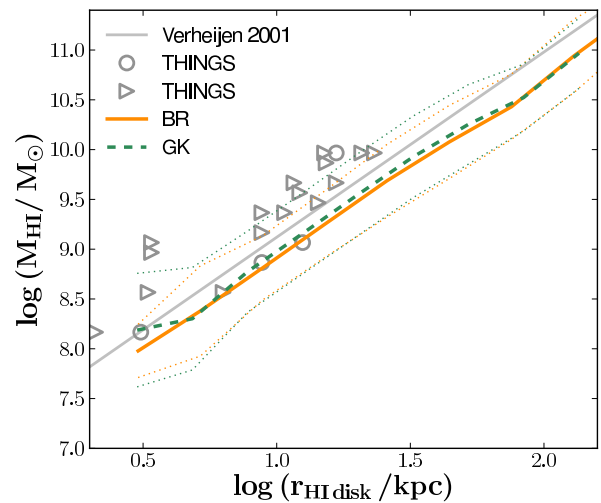}
\caption{The `\hi radius', defined as the radius at which
  $\Sigma_{\rm{HI}} = 1\,M_\odot\,\rm{pc}^{-2}$ as a function of \hi
  mass for the pressure- (orange solid) and metallicity-based (green
  dashed) \h2 formation recipes. Thick lines show the mean,
    and dotted lines mark the $2\sigma$ deviation. The grey
  circles/arrows are measurements/lower limits obtained from the
  profiles presented in \citet{Leroy2008}. The solid grey line is a
  fit to the data presented in
  \citet{Verheijen2001}.  \label{fig:HI_size}}
\end{figure}

\begin{figure*}
\includegraphics[width = 0.9\hsize]{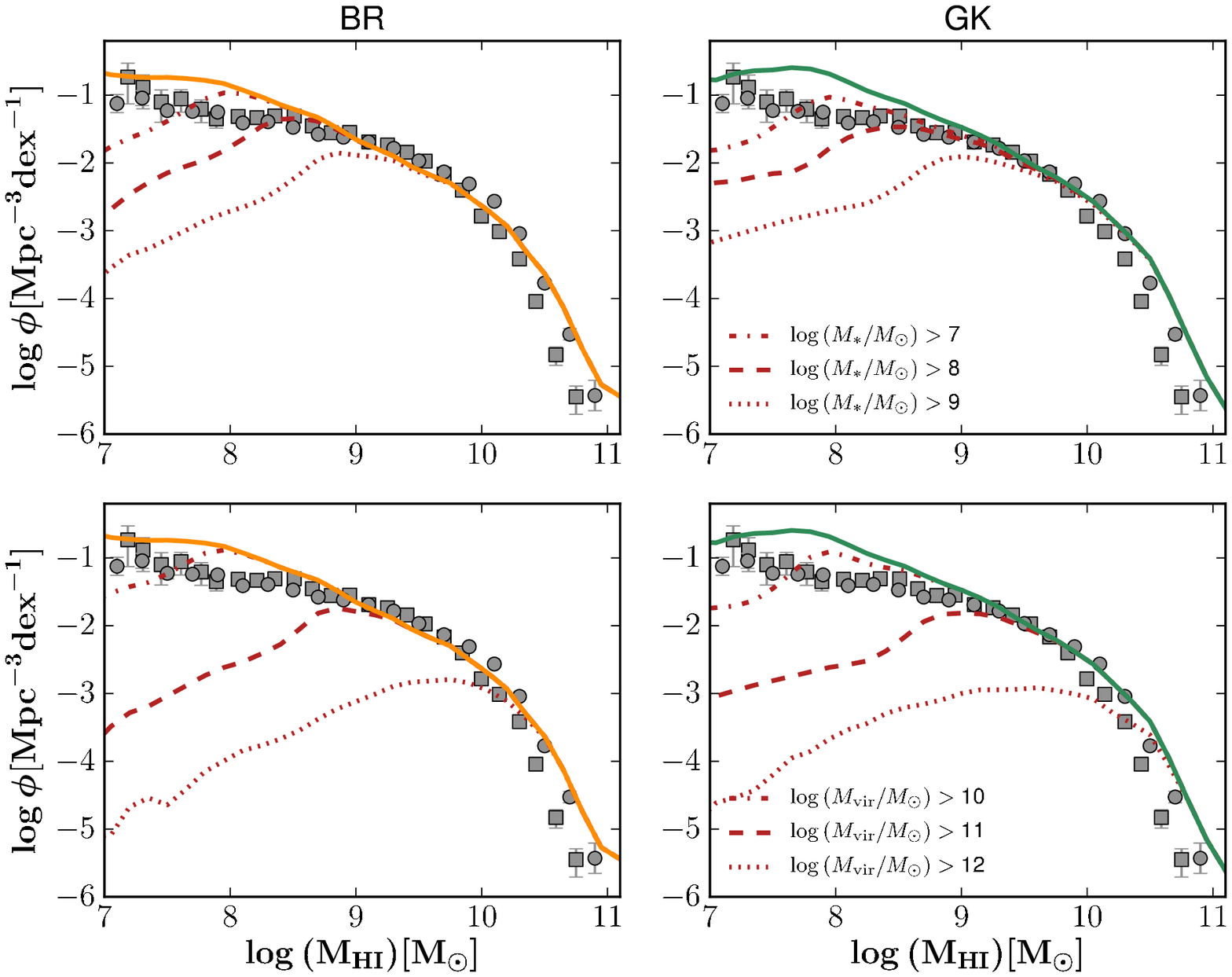}
\caption{The predicted \hi mass function of galaxies at $z=0.0$, assuming a
  pressure- (left) and metallicity-based (right) \h2 recipe. The top
  row shows the contribution to the mass function by galaxies with a
  lower cutoff in stellar mass. The bottom row shows the contribution
  by galaxies hosted by halos with a cutoff in virial mass. Grey
  circles and squares show the observed $z=0$ \hi mass functions from
  \citet{Zwaan2005} and \citet{Martin2010},
  respectively.\label{fig:HI_mass_func_z0.0}}
\end{figure*}

\begin{figure*}
\includegraphics[width = 0.9\hsize]{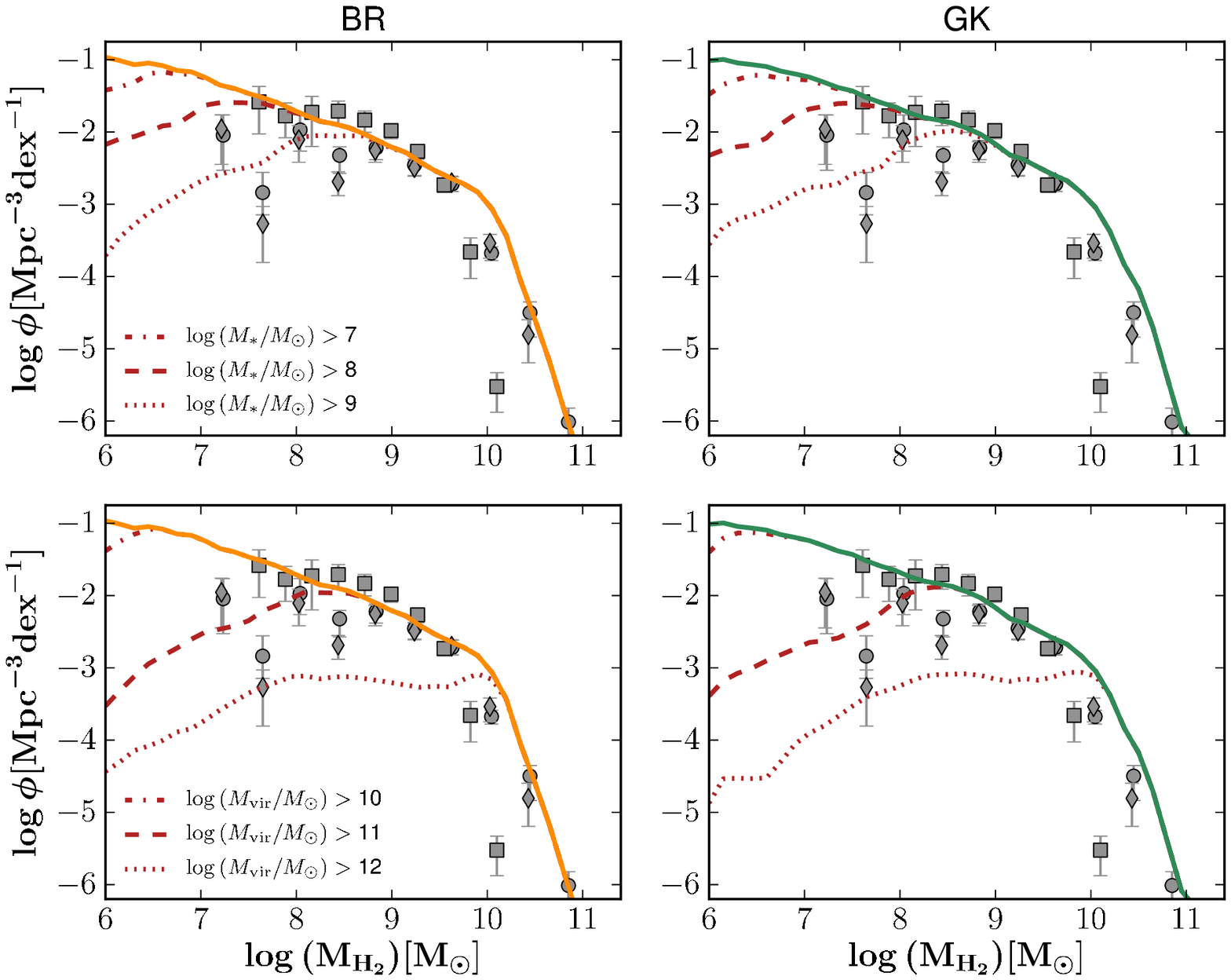}
\caption{The predicted \h2 mass function of galaxies at $z=0.0$,
  assuming a pressure- (left) and metallicity-based (right) \h2
  recipe. The top row shows the contribution to the mass function by
  galaxies with a lower cutoff in stellar mass. The bottom row shows
  the contribution from galaxies hosted by a halo with a lower cutoff
  in virial mass. Grey circles and diamonds are the estimated observed $z=0$ \h2
  mass functions from \citet{Keres2003}, grey squares are the
  observational estimates presented by \citet{Obreschkow2009} (see
  text). \label{fig:H2_mass_func_z0.0}}
\end{figure*}
\begin{figure*}
\includegraphics[width = 0.9\hsize]{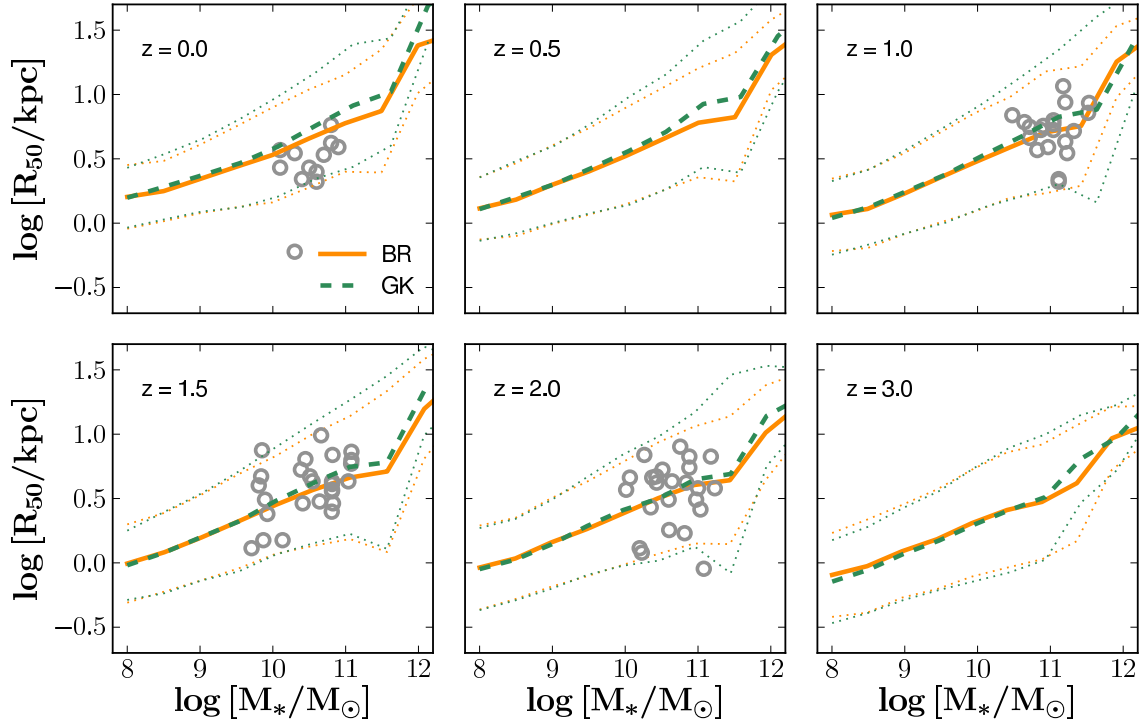}
\caption{The SFR half-light radii of galaxies as a function of their
  stellar mass for different redshift bins, for the pressure-based
  (orange solid) and metallicity-based (green dashed) models. Thick lines show the mean of the distribution, and dotted lines
    mark the $2\sigma$ deviation. Grey circles are observations from
  \citet[at $z=0.0$]{Leroy2008}, \citet[H$_{\alpha}$ half light
    radii]{Foerster2009}, \citet{Genzel2010} and
  \citet{Tacconi2013}.\label{fig:CO_size}}
\end{figure*}

\begin{figure*}
\includegraphics[width = 0.9\hsize]{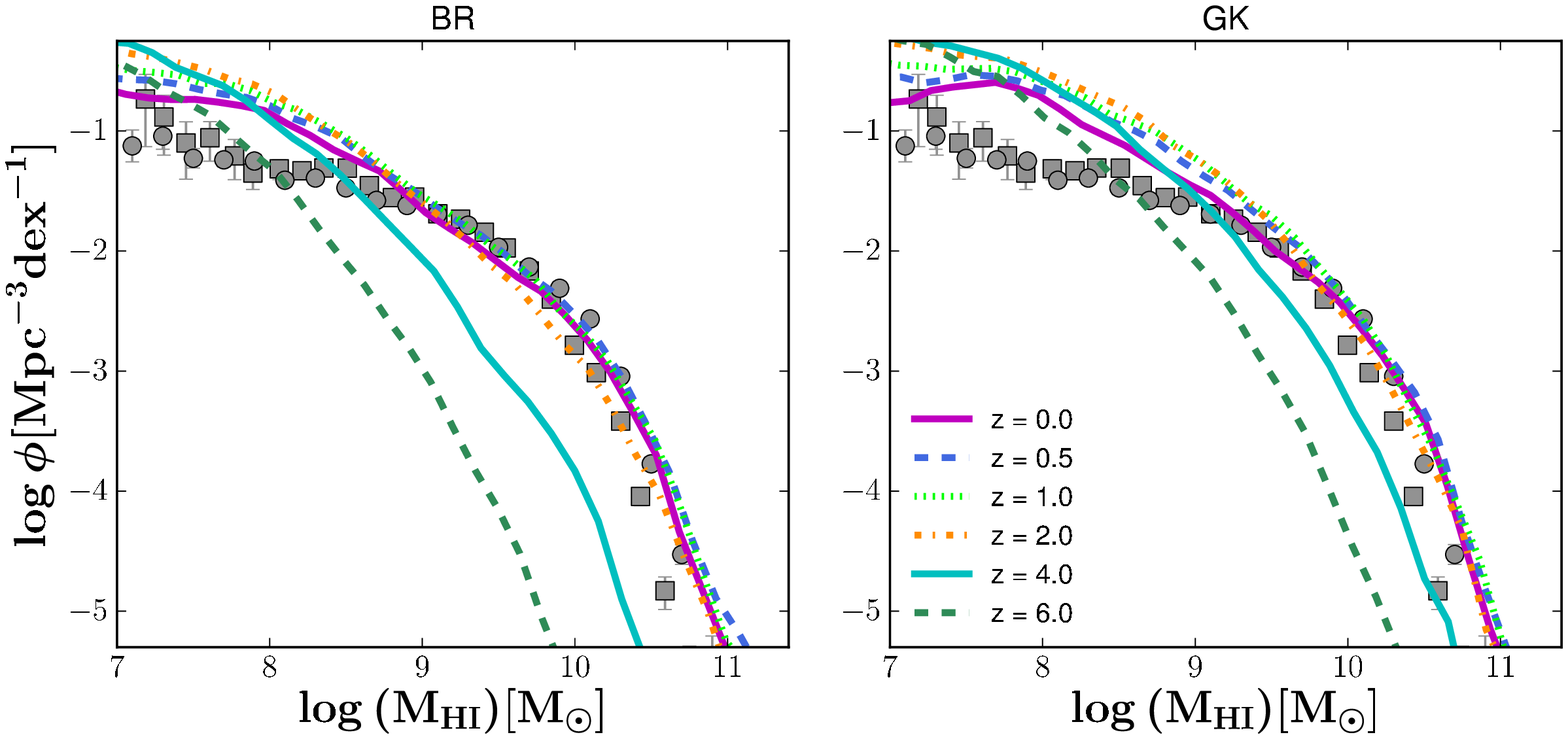}
\caption{Predicted redshift evolution of the \hi mass function,
  assuming a pressure- (left) and metallicity-based (right) \h2
  recipe. Grey circles and squares show the observed $z=0$ mass
  functions from \citet{Zwaan2005} and \citet{Martin2010},
  respectively.\label{fig:HI_mass_func_evol}}
\end{figure*}
\subsubsection{Molecular based star formation}
Star formation is modeled following empirical relationships from
recent observations. \citet{Bigiel2008} suggest, based on observations
of spiral galaxies from the THINGS survey, that the star-formation
rate surface density can be directly related to the surface density of
molecular gas, i.e.
\begin{equation}
\label{eqn:bigiel1}
 \Sigma_{\rm SFR} = A_{\rm SF} \, {\Sigma_{\rm H_2}}^{N}
\end{equation}
with $N\simeq 1$. Observations of higher density
environments suggest that above some critical \h2 surface density, the
slope of the relation described in equation \ref{eqn:bigiel1}
steepens. We therefore adopt a two-part scaling law given by:
\begin{equation}
\label{eqn:bigiel2}
\Sigma_{\rm SFR} = A_{\rm SF} \, (\Sigma_{\rm H_2}/10 M_\odot {\rm pc}^{-2}) \left(1+
\frac{\Sigma_{H_2}}{\Sigma_{\rm H_2, crit}}\right)^{N_{\rm SF}}
\end{equation}
We adopt the ``two-slope'' star formation recipe in all of
  the models presented in this work. In addition, we adopt $A_{\rm
    SF}=5.98 \times 10^{-3}\, M_\odot {\rm yr}^{-1} {\rm kpc}^{-2}$,
  $\Sigma_{\rm H_2, crit} = 70 M_\odot$ pc$^{-2}$, and $N_{\rm
    SF}=1.0$. The value of $A_{\rm SF}$ is taken from the observations
  of \citet{Bigiel2008}, corrected to our system in which Helium is
  not included in the gas masses and densities. The values for
  $\Sigma_{\rm H_2, crit}$ and $N_{\rm SF}=1.0$ are motivated by the
  observational compilation presented in \citet{Narayanan2012}.
\begin{figure*}
\includegraphics[width = 0.9\hsize]{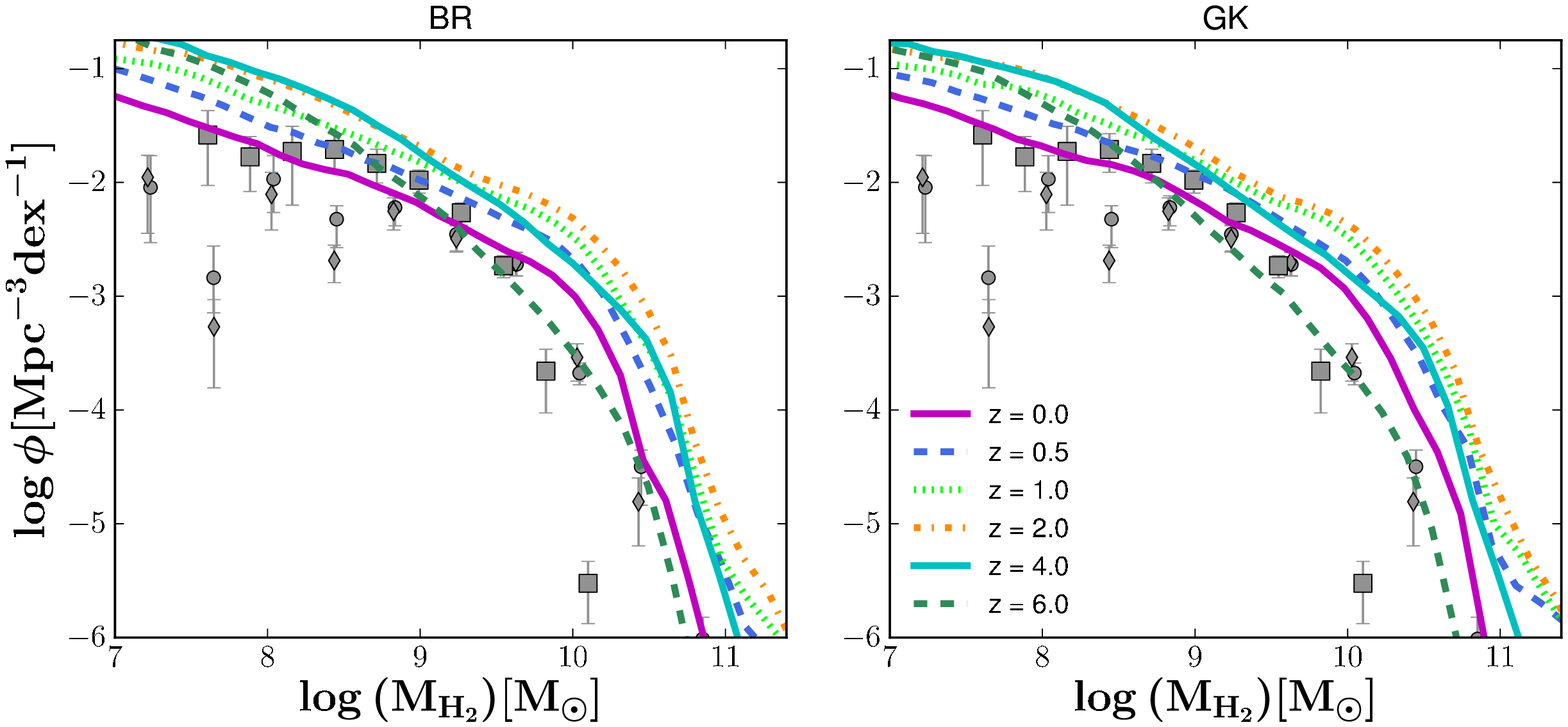}
\caption{Predicted redshift evolution of the \h2 mass function,
  assuming a pressure- (left) and metallicity-based (right) \h2
  recipe. Grey circles are the $z=0$ observational estimates of the
  \h2 mass functions from \citet{Keres2003}, and grey squares are the
  estimates from
  \citet{Obreschkow2009}. \label{fig:H2_mass_func_evol}}
\end{figure*}

\begin{figure*}
\includegraphics[width = 1.0\hsize]{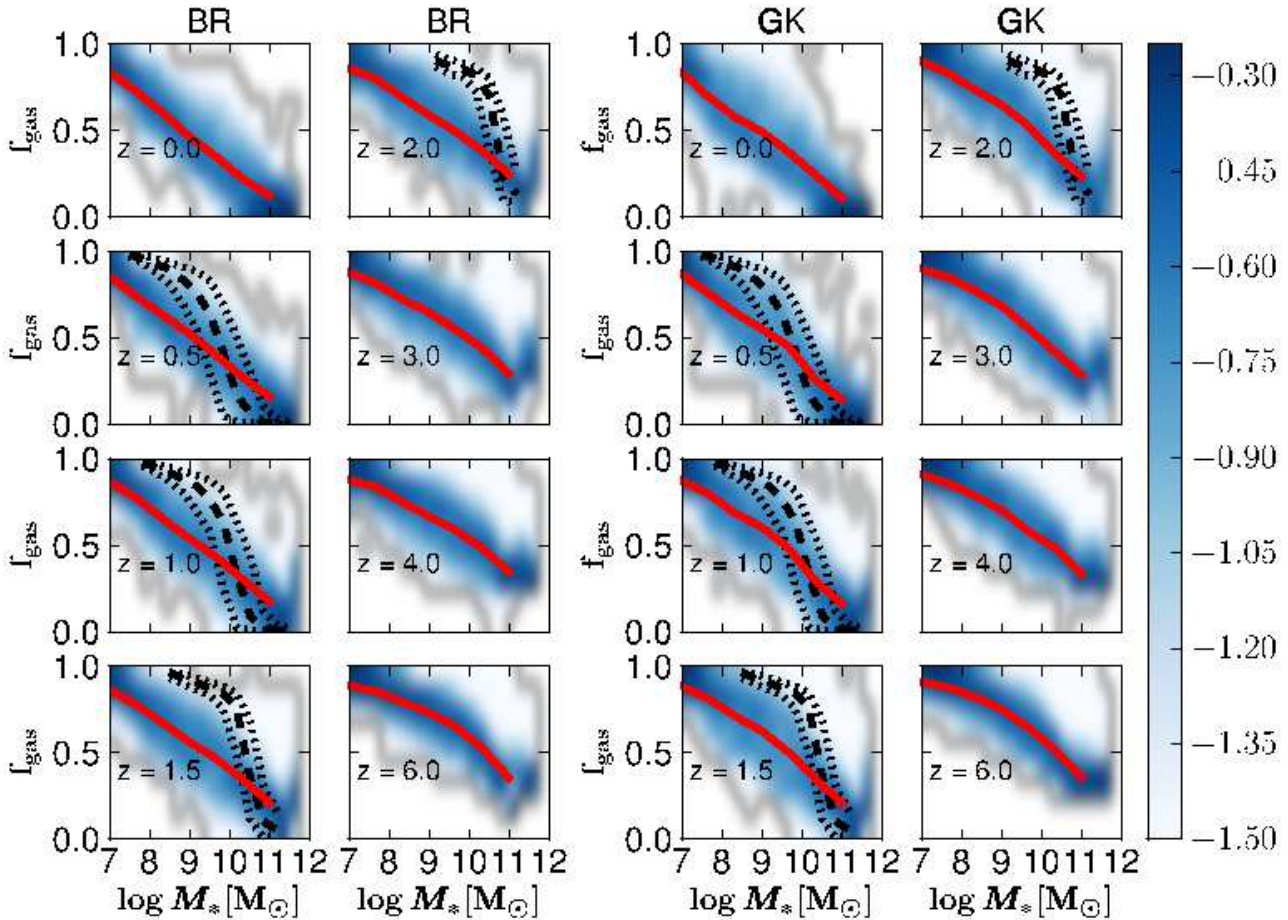}
\caption{Galaxy cold gas fraction ($f_{\rm{gas}} \equiv
\frac{M_{\rm{H2 + HI}}}{M_{\rm{H2 + HI}} + M_*}$) as a function of stellar mass in
  disc-dominated galaxies for different redshift bins for the
  pressure-based \h2 prescription (left column) and the
  metallicity-based prescription (right column). The blue shaded
  regions show the log of the conditional probability distribution
  function $P(f_{\rm{gas}}|M_*)$, whereas the red solid lines shows
  the median fit. The black dashed
  and dotted lines show the mean and two sigma confidence region from
  indirect observational estimates of the gas fraction from
  \citet[][see text for details]{Popping2012}.
\label{fig:gas_frac_evol}} 
\end{figure*}

\begin{figure*}
\includegraphics[width = 1.0\hsize]{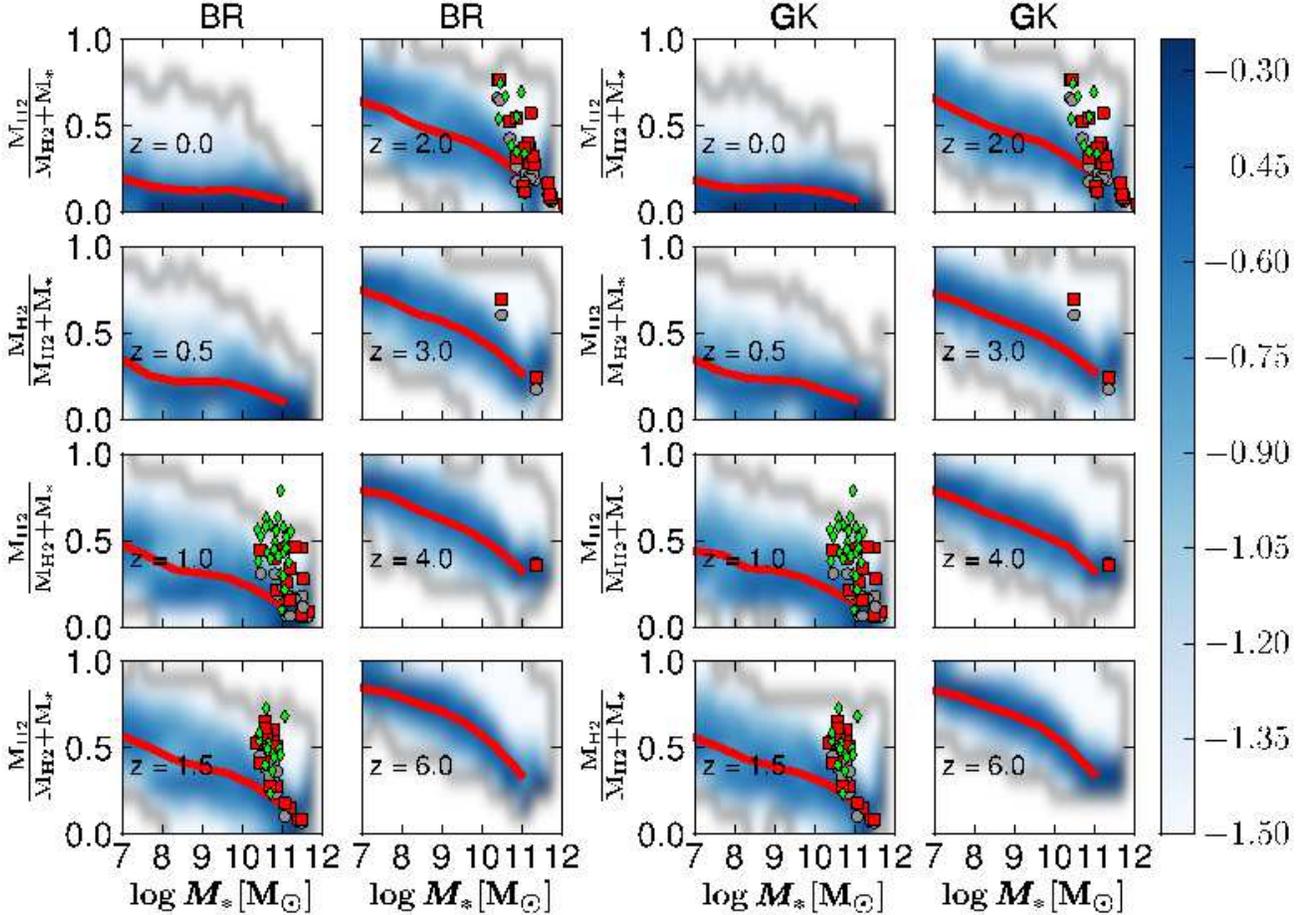}
\caption{Molecular fraction 
$\frac{M_{\rm{H2}}}{M_{\rm{H2}} + M_*}$ as a function
  of stellar mass in disc-dominated galaxies for different redshift
  bins for the pressure-based \h2 prescription (left column) and the
  metallicity-based prescription (right column). Blue shading shows the
  log of the conditional probability distribution function
$P(\frac{M_{\rm{H2}}}{M_{\rm{H2}} + M_*}|M_*)$, whereas the red solid line shows
  the median fit. Grey circles
  and red squares are estimates taken from
  \citet{Narayanan2012_gasfrac} using the traditional and newly
  calculated value for the CO-to-H2 conversion, respectively. Green
  diamonds are observations from \citet{Tacconi2013}. The black dashed
  and dotted lines show the mean and two sigma confidence region of
  the gas fractions presented in \citet{Popping2012}.\label{fig:H2_star_frac_evol}}
\end{figure*}

\begin{figure*}
\includegraphics[width = 1.0\hsize]{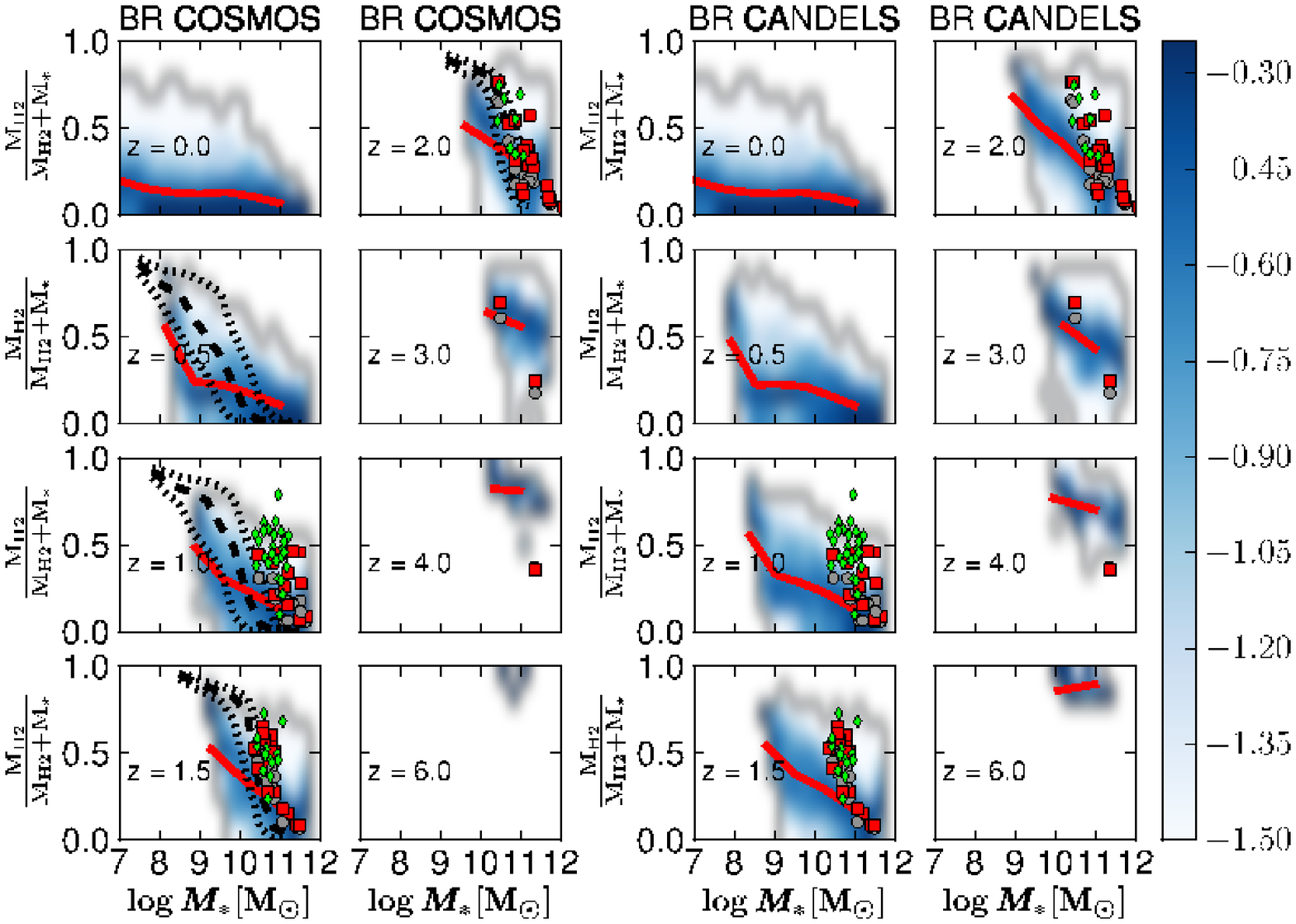}
\caption{Molecular fraction $\frac{M_{\rm{H2}}}{M_{\rm{H2}} + M_*}$ as a function
  of stellar mass for different redshift bins, assuming a metallicity-based star formation law. We applied the selection criteria for
  \citet[][{\it COSMOS}, I$_{\rm{AB}}<$24; left column]{Popping2012}
  and {\it CANDELS} (H$_{\rm{AB}}<$ 25; right column). Blue shading
  shows the log of the conditional probability distribution function
  $P(\frac{M_{\rm{H2}}}{M_{\rm{H2}} + M_*}|M_*)$, whereas the red solid line shows
  the median fit. The dashed
  and dotted lines represent the mean and two sigma confidence region
  of the gas fractions presented in \citet{Popping2012}. Grey circles, red squares, and
  green diamonds are as in Figure
  \ref{fig:H2_star_frac_evol}.\label{fig:frac_evol_literature_GK}}
\end{figure*}
\begin{figure}
\includegraphics[width = \hsize]{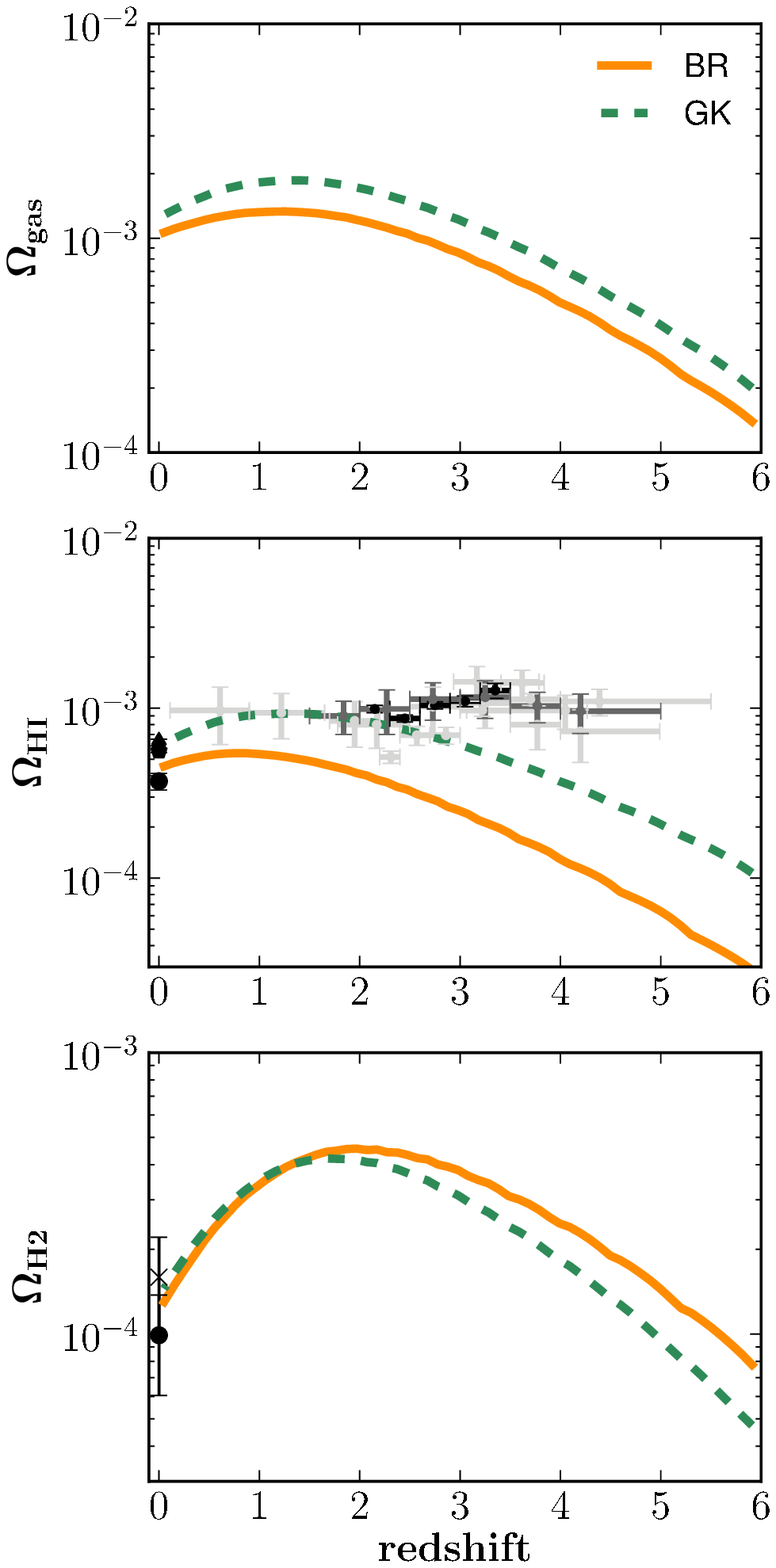}
\caption{The cosmic comoving density, in units of the critical
  density, of cold gas (\hi + \h2 + \hii; top panel), \hi (middle) and \h2 (bottom) as a
  function of redshift. The solid orange line shows the pressure-based
  and the green dashed line shows the metallicity-based \h2
  recipes. Observations of \citet{Peroux2005,Rao2006,Guimaraes2009,Prochaska2009} are
  overplotted in light gray. Dark gray observations are by
  \citet{Zafar2013} and observations from \citet{Noterdaeme2012} and
  local galaxies \citep{Keres2003,Zwaan2005,Martin2010,Obreschkow2009,Braun2012} are
  overplotted in black.\label{fig:density_evolution}}
\end{figure}
\begin{figure}
\begin{center}
\includegraphics[width = 0.9\hsize]{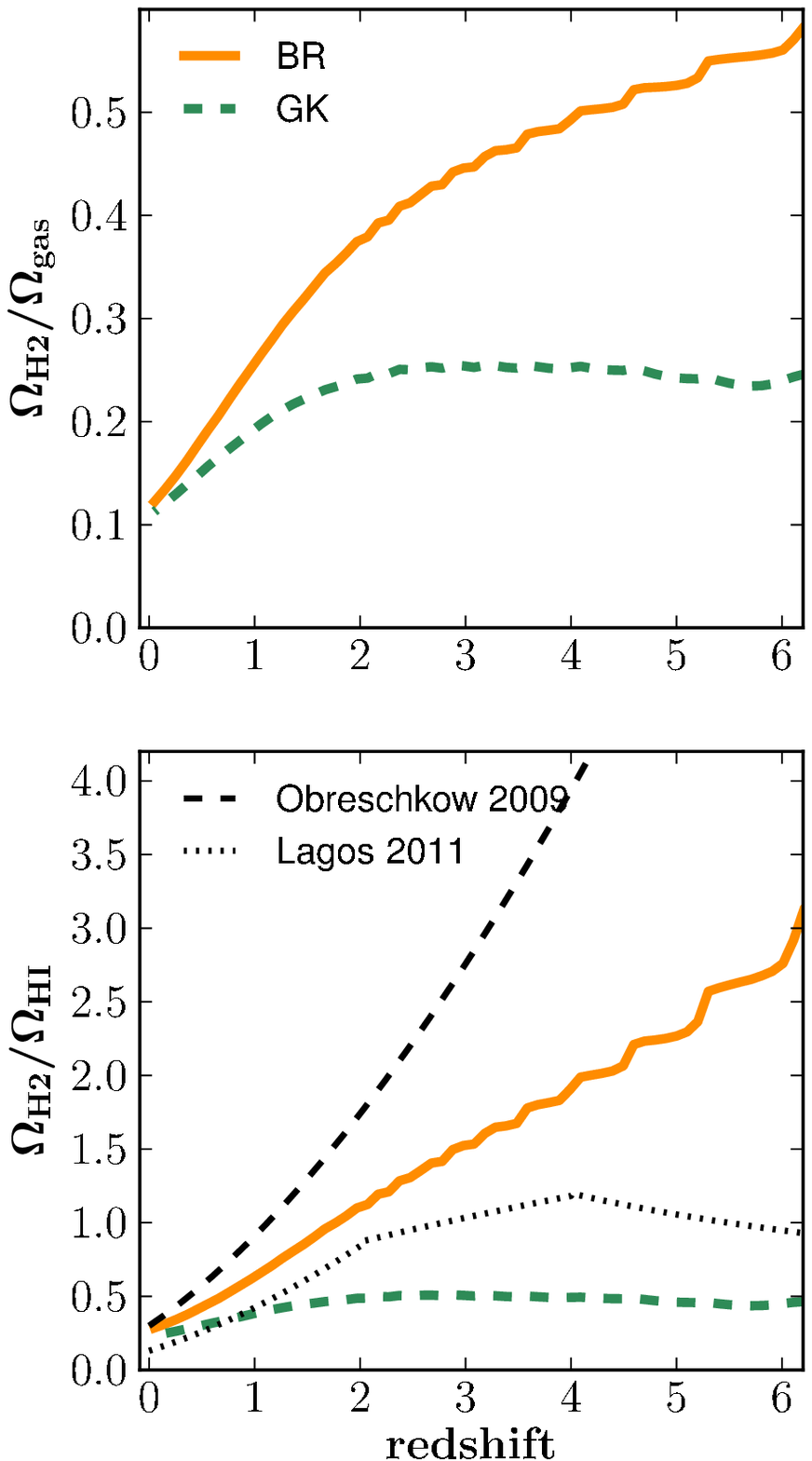}
\caption{The cosmic \h2 fraction (top panel) and \h2-to-\hi ratio
  (bottom panel) as a function of redshift for the pressure- (solid orange) and metallicity-based (green dashed) \h2 formation recipes. Pressure- and
  metallicity-based \h2 recipes are marked with solid orange and green
  dashed lines, respectively.\label{fig:density_ratio_evolution}}
\end{center}
\end{figure}  

\section{Results}
\label{sec:results}
In this section we show our predictions for the evolution of the \hi
and \h2 content of galaxies over a range of redshifts from $z=0.0$ to
$z = 6.0$. The simulations were run on a grid of halos with virial
masses ranging from $5\times 10^8 M_\odot$ to $5\times 10^{14}
M_\odot$ with a mass resolution of $5\times 10^6 M_\odot$.
We first perform a comparison of our model predictions with
observations of local galaxy properties, in order to validate our
models. All presented gas masses are pure hydrogen masses and do not
include a correction for Helium.

\subsection{Local galaxy properties}
In Figure \ref{fig:gas_scaling} we present the ratios of \hi and \h2
relative to stellar mass, and the ratio of \h2 to \hi, as functions of
stellar mass and stellar surface density in disc-dominated galaxies ($M_{*_{\rm{bulge}}}/M_{*_{\rm{total}}} \leq 0.4$). We compare our results to a
compilation of observations presented in \citet{Leroy2008},
\citet{Saintonge2011}, and \citet{Catinella2013} based on the THINGS+HERACLES and GASS+COLDGASS
surveys and in \citet{Boselli2014} based on the Herschel reference survey.

Both the pressure-based and metallicity-based recipes show very good
agreement with the observed trends between stellar mass or stellar
surface density and \hi and \h2 fractions. The fraction of \hi
relative to stars decreases with increasing stellar mass and surface
density, whereas the fraction of \h2 relative to stars remains roughly
constant. Consequently, the fraction of cold gas in the form of \h2
increases with stellar mass and surface density. The \h2-to-\hi ratio
as a function of stellar mass is on average slightly too high in our
models, although still within the scatter of the observations
(particularly at low stellar masses). Here we focus on the gas
  fractions of disc-dominated galaxies. A similar exercise for a
  ``blind'' survey of galaxies would yield lower \hi-to-stellar mass,
  \h2-to-stellar mass, and \h2-to-\hi mass ratios. Spheroidal objects
  have much lower relative gas content than disc-dominated galaxies
  and most of the cold gas is atomic.

We present the ratio of ionised hydrogen to galaxy stellar mass as a
function of stellar mass in Figure \ref{fig:HII}. We find a
monotonic decrease in the ratio between ionised hydrogen and stellar
mass, without any significant difference between the two \h2 formation
recipes. The ionised hydrogen mass ranges from about a tenth of the stellar
mass in large galaxies, to about equal to the stellar mass in low mass
objects ($\log{(M_*/M_\odot)} < 8 \,-\,9$), and up to 10-100 times the
stellar mass in very low mass galaxies ($\log{(M_*/M_\odot)} < 6 \,-\,7$). These ratios are comparable to the ratio between galaxy \hi
and stellar mass, indicating that a significant amount of the cold gas
in galaxies may be in an ionised component, as suggested by recent
observations \citep{Tumlinson2011}, but not accounted for in previous
semi-analytic models of galaxy formation. We will further explore the
predictions for ionised hydrogen in galaxies in future work.

Both \h2 recipes presented in this work rely on the estimated size of
the galaxy disc, as this sets the surface density of the gas, one of
the key parameters in calculating the \h2 content of the gas. It is
therefore of great importance to correctly predict the sizes of the
gas disc in galaxies. Figure \ref{fig:HI_size} shows the \hi disc size
of a galaxy as a function of its \hi mass. Following
\citet{Verheijen2001}, we define the size of the \hi disc as the
radius at which the \hi surface density of the gas equals
$\Sigma_{\rm{HI}} = 1 M_\odot\,\rm{pc}^{-2}$. We calculate the
location of $\Sigma_{\rm{HI}} = 1 M_\odot\,\rm{pc}^{-2}$ in
post-processing, assuming an exponential distribution of the cold gas
in the disc. Besides the fit presented in \citet{Verheijen2001}, we
also present, where possible, the size of the \hi discs of the THINGS
sample of galaxies, which we computed from the radial profiles
presented in \citep{Leroy2008}.

There is good agreement between the modeled and observed \hi disc
sizes, spanning over three orders of magnitude in \hi mass and two
orders of magnitude in disc size. The good agreement between model and
data is independent of the \h2 formation prescription.  
We have shown that our models match the observed \h2 fractions
  for nearby galaxies (Fig. \ref{fig:gas_scaling}), and
  \citet{Somerville2008size} has shown previously that the models also
  reproduce the size-stellar mass relation for disc-dominated galaxies
  from $z\sim 2$ to the present.  Although this does not necessarily
guarantee a match between the \hi disc size and gas content of a
galaxy, the agreement is an encouraging sanity check.

Figure \ref{fig:HI_mass_func_z0.0} shows our predictions for the \hi
mass-functions at $z = 0.0$. Both star formation recipes show decent
agreement with the observed \hi mass functions at \hi masses of
$\log{(\rm{M}_{\rm{HI}}/\rm{M}_\odot)} \sim 10$ and higher. The
pressure-based recipe slightly underpredicts the observed \hi mass
function in the mass range $\log{(\rm{M}_{\rm{HI}}/\rm{M}_\odot)} \sim
9 -10$, and slightly overpredicts the observations at lower \hi
masses. The metallicity-based recipe overpredicts the observed number
of galaxies below $\log{(\rm{M}_{\rm{HI}}/\rm{M}_\odot)} \sim
8.5$. 

Figure \ref{fig:HI_mass_func_z0.0} shows that the galaxies responsible
for the excess of low-\hi mass objects are low-mass galaxies
($\log{(M_*/M_\odot)} \leq 7$) residing in low mass halos
($\log{(M_{\rm{vir}}/M_\odot)} \leq 10$). This underlines the
  importance of sufficiently high mass resolution in simulations that
  attempt to predict the properties of galaxies observed in \hi.

The predicted \h2 mass function at $z=0.0$ is presented in Figure
\ref{fig:H2_mass_func_z0.0}, and compared with two observational
estimates. Both estimates are based on the CO survey of
\citet{Keres2003}. The estimated \h2 mass function given by
\citet{Keres2003} was obtained by applying a fixed conversion factor
to convert between CO and \h2. \citet{Obreschkow2009} estimated a
variable \h2-CO conversion factor based on the galaxy
properties. Based on recent observations and theoretical work, a
variable conversion factor that depends on galaxy properties (such as
metallicity) is probably more reasonable (we discuss this further
below). The predictions of both recipes are very similar, and we
obtain good agreement with the observational estimates of
\citet{Keres2003}, but significantly overproduce galaxies with large
\h2 masses relative to the \citet{Obreschkow2009} results. It is
possible that a process not included in our model, such as AGN
feedback, could destroy or expel \h2 in massive galaxies
\citep{Saintonge2012} and possibly lower the number of \h2
massive galaxies.

\subsection{Evolution of gas in galaxies}
In this section we present our predictions for the evolution of the
gas content in galaxies and make predictions for upcoming surveys of
gas at high redshifts. 

\subsubsection{Galaxy sizes}
Figure \ref{fig:CO_size} shows the SFR half-light radius of our
modelled galaxies as a function of their stellar mass (i.e., the
radius that encompasses half of the total SFR of the galaxy). We
compare these results with radii presented in the literature for
high-redshift galaxies \citep{foerster2006,Genzel2010,Tacconi2013}
and, where possible, the CO half-light radius of the discs of the THINGS
galaxies, which we computed from the radial profiles presented in
\citep[][assuming a fixed conversion between the \h2 and CO radial profiles]{Leroy2008}. Our results are in
excellent agreement with the observations at high-redshift and in the
local Universe, indicating that in spite of
the simplicity of our model for computing disc sizes and surface
density profiles, we appear to be able to correctly model the sizes and the
location of star formation and the evolution of these quantities since
$z=2$. For a fixed stellar mass, the SFR half-light radius increases
with decreasing redshift. Consequently, the molecular gas is more
compact in high redshift galaxies. 
This behavior is driven by the overall growth of galaxy discs with time, as
they accrete gas with higher angular momentum.

\subsubsection{\hi mass function}
Figure \ref{fig:HI_mass_func_evol} shows the predicted \hi mass
function at redshifts between $z=0$ and $z=6$. We overplot
observations from \citet{Zwaan2005} and \citet{Martin2010} at $z=0.0$.
For \hi masses ($\log{(M_{\rm{HI}}/M_\odot)} \gtrsim 8$), the figure
shows a clear monotonic increase in the number of galaxies at a given
\hi mass from $z=6$ to $z=2.0$. There is very weak evolution at
$z\lesssim 2$, and almost none at all from $z\sim 1$--0.  The weak
evolution in the number of low \hi mass galaxies shows that in our
current model framework, the excess of these objects is already
present at redshifts $\sim$ 2. We find little difference in the predicted evolution
of the \hi mass function between the metallicity- and pressure-based
recipes. 

Although little evolution is seen in the \hi mass function since
$z\sim 2$, this of course does not mean that galaxies are static, or
that \hi is not being created or destroyed. It rather means that there
is a kind of self-regulated equilibrium that arises naturally in these
models. 

\subsubsection{\h2 mass function}
In Figure \ref{fig:H2_mass_func_evol} we show the predicted \h2 mass
function at redshifts between $z=0$ and $z=6$. The left panel contains
mass functions obtained using the metallicity-based recipe, whereas
the right pannel shows results obtained using a pressure-based
recipe. These predictions are compared with the observational estimates of the
\h2 mass function at $z=0$, as shown in
Figure~\ref{fig:H2_mass_func_z0.0}.

Both \h2 recipes predict a gentle evolution in the \h2 mass function
at all \h2 masses. In both recipes, the number of galaxies with large
\h2 masses increases from $z\sim 6$--2, then declines slightly to
$z=0$. At lower masses, $\log{(M_{\rm{H2}/M_\odot})} \lesssim 9$, both
models predict a slight increase in the number of low-\h2 mass
galaxies from $z\sim 6$--4, then a more or less monotonic
decline from $z\sim 4$ to $z\sim 0$. 

In both recipes, it is more difficult to form \h2 in low-surface
density gas. In our models, low-mass halos host galaxies with a larger
fraction of their gas at low surface density (this is in accord with
observational size-mass scaling relations), and therefore low-mass
galaxies are less efficient at forming \h2, as we saw in
Figure~\ref{fig:gas_scaling}. In the BR model, we would say this is
because their disc midplane pressure is lower due to their smaller
gravitational potential wells. In the GK model, we would say it is due
to the lower availability of dust grains on which \h2 can form. Thus
the build-up of large \h2-mass galaxies from $z\sim 6$--2 reflects the
growth of structure and the formation of massive dark matter halos,
while the decrease in the number of low-\h2 mass galaxies from
$z\sim 4$ to $z\sim 0$ reflects the growth of galaxy discs resulting
in lower cold gas surface densities, combined with low potential
wells and/or low availability of dust grains. 

\subsubsection{Evolution in galaxy gas-fractions}
In the following figures we present the gas fraction and relative \h2
content of galaxies as a function of their stellar mass
for different redshifts ($0<z<6$). In each case, we plot the
conditional probability $P(f_{\rm gas}|m_{\rm star})$, and the reader
should keep in mind that the most massive galaxies will be extremely
rare at high redshift, and probably would not be included in any
observed samples.

Figure \ref{fig:gas_frac_evol} shows the cold gas fraction of the
modeled galaxies as a function of stellar mass, divided into redshift
bins. We also included the indirectly derived gas fraction from
\citet{Popping2012}. They calculated cold gas and \h2 masses in
galaxies from the {\it COSMOS} survey by inverting the
\citet{Bigiel2008} star-formation law in combination with the
\citet{Blitz2006} method to calculate the \h2 fraction of cold
gas. Including a recipe to calculate the \h2 fraction of cold
  gas allowed Popping et al. to indirectly estimate both the molecular
  and the atomic hydrogen masses of thse galaxies.

Our models predict that gas fractions decrease only mildly
from $z\sim 6$--3. At lower redshifts the gas fractions decrease
rapidly, such that galaxies with large stellar masses run out of gas
first. This evolution is similar for both \h2 recipes. Only in low
mass galaxies ($\log{(M_*/M_\odot)} \leq 9$) do the two applied
recipes give different predictions, with the metallicity-based recipe
predicting slightly larger gas fractions. We find that our model
predictions are in good agreement with the indirect estimates of
\citet{Popping2012} for $z\leq1.0$ in the mass range
$\log{(M_*/M_\odot)} > 10$. At higher redshifts we find good agreement
for objects with $\log{(M_*/M_\odot)} \geq 10.5$. We overpredict the
indirect estimates from the literature at lower stellar mass, however,
we did not take the selection criteria applied by Popping et al. into
account here.

Figure \ref{fig:H2_star_frac_evol} shows $f_{H2} \equiv
\frac{M_{\rm{H2}}}{M_{\rm{H2}} + M_*}$ as a function of stellar mass
at different redshifts.  We included a compilation of observations
presented in \citet[taken from
\citet{Genzel2010,Daddi2010,Tacconi2010,Casey2011,Bothwell2013}]{Narayanan2012_gasfrac}
and in \citet{Tacconi2013}. Besides
the \h2 masses quoted in the original literature,
\citet{Narayanan2012_gasfrac} uses a novel approach to calculate the
conversion between CO observation and \h2 masses and their resulting
gas fractions (see section \ref{sec:CO} for a detailed
description). We included the original values for
$f_{H2}$ as well as the recalibrated values. Similar to the previous
figure, there is no significant difference between the two studied
recipes. The evolution in $f_{H2}$, however, is much stronger. At $z
=0.0$ we find $f_{H2} \sim0.1$ at all probed stellar masses, whereas
at $z=6.0$ we find values of $f_{H2} \sim 0.8$ over a large range of
stellar masses. There is large scatter in $f_{H2}$ at redshifts $z=3.0
- 0.5$ over all probed stellar masses. This scatter is indicative of a
transitional phase during which the relative \h2 content of galaxies
rapidly drops, however, this does not necessarily take place at the
same time/rate in galaxies with similar stellar mass.

This strong evolutionary trend, compared to the trends seen for the
total cold gas fraction, indicates that the amount of \h2 decreases
not only due to the availability of less cold gas, but that the \h2
fraction itself also drops \citep{Popping2012}. The rate at which this
happens is independent of adopted recipe in our models. We find good
agreement with the observations and their re-analysis by
\citet{Narayanan2012_gasfrac}. Our model does not strongly favor
either choice for the CO-\h2\ conversion factor.  Similar to the
total cold gas fractions, we find that our model predicts a lower
relative \h2 content of galaxies than the indirect estimates by
\citet{Popping2012} suggest (especially at stellar masses
$\log{(M_*/M_\odot)} \leq 10.5$). We again emphasize that so far, we
did not take the selection bias inherent to the observations that went
into \citet{Popping2012} analysis into account. We will now discuss
how selection criteria affect our results.

Current samples of high-redshift galaxies are highly sensitive to
their selection criteria and direct observation of the molecular
content of the galaxies are usually biased towards the most gas rich
galaxies. To study how this bias might affect the comparison of our
model predictions with observations in the literature, we apply the
selection criteria from the relevant surveys to our model galaxies
assuming a metallicity-based \h2 recipe and show the results in
Figure. \ref{fig:frac_evol_literature_GK}. We compare our results to
the gas fraction estimates for galaxies taken from the {\it COSMOS}
sample with $I_{\rm{AB}} < 24$ \citep{Popping2012}. We also show
predictions for a sample with H$_{\rm{AB}}<$ 25 mag, representative
of galaxies in the CANDELS survey \citep{Grogin2011,Koekemoer2011} for which reliable measurements of
galaxy size are expected to be able to be obtained.

When we account for the selection effects, we find good agreement with
the indirect \h2 fraction estimates from the {\it COSMOS} sample. At $z>1.5$ our model predicts slightly lower gas fractions than
  those suggested by the indirect estimates. The rough agreement is a
very encouraging result for our model, but also emphasizes how
important it is to properly take selection criteria into account when
comparing models to observed galaxy samples. Our results also suggest
that repeating the analysis on the deeper, $H$-band selected CANDELS
sample will greatly expand the range of stellar mass and gas fraction
that can be probed by the indirect method at $z>1.5$. We intend to
repeat the \citet{Popping2012} analysis on the CANDELS sample in the
near future. These results will provide an interesting complement to
direct measures of high redshift gas fractions that will become
available from ALMA.

\subsubsection{Gas density evolution of the Universe}
Figure \ref{fig:density_evolution} shows the predicted global \hi,
\h2, and total cold gas density (including ionised hydrogen) of the Universe as a function of time (in
units of the critical density). We compare our results to local \hi
and \h2 densities
\citep{Keres2003,Zwaan2005,Obreschkow2009,Martin2010,Braun2012} and
high-z estimates of the HI density obtained from Damped Lyman-$\alpha$
(DLA) absorption systems
\citep[e.g.,][]{Peroux2005, Rao2006,Guimaraes2009,Prochaska2009,Noterdaeme2012,Zafar2013}.

We see that the two \h2\ formation recipes differ significantly in
terms of both the total cold gas content of the Universe and the ratio
between \hi\ and \h2. The metallicity-based recipe predicts more cold
gas overall at all redshifts, and also more \hi. The pressure-based
recipe produces more \h2\ overall, in spite of the lower amount of
total cold gas. Both models underpredict $\Omega_{\rm HI}$ inferred
from DLAS at $z\geq3$, the pressure-based model more dramatically. On
the other hand, predictions by the metallicity-based model is in
decent agreement with DLA observations at $z<2.5$. Overall the
metallicity-based recipe is better in reproducing the observed values
for $\Omega_{\rm HI}$ and $\Omega_{\rm H2}$.

The two \h2\ formation recipes show a very different evolution in the
global ratio of \hi\ to \h2 with redshift (see Figure
\ref{fig:density_ratio_evolution}). The pressure-based recipe predicts
a monotonic increase in $\Omega_{H2}/\Omega_{\rm gas}$ and
$\Omega_{H2}/\Omega_{HI}$ with increasing redshift. The
metallicity-based recipe predicts a very mild increase with increasing
redshift up to $z>3.0$, then a flattening at higher redshifts.
Especially worthwhile to note is that $\Omega_{H2}$ never exceeds
$\Omega_{HI}$ for the metallicity-based recipe, whereas it does by up
to a factor of three for the pressure-based recipe. We will give a
detailed discussion about the origin of these differences, and how
they can help to constrain the physics driving the partitioning of
hydrogen into atomic and molecular hydrogen, in Section
\ref{sec:discussion}. 

As a comparison we also show predictions from
  \citet{Obreschkow2009_letter} and \citet{Lagos2011cosmic_evol} in
  Figure \ref{fig:density_ratio_evolution}. Both authors use a
  pressure-based recipe similar to ours. Although the predictions
  differ in detail --- unsurprising as many other aspects of the
  models differ --- we find that our predictions for the
  pressure-based recipe are in qualitative agreement with other
  predictions from the literature, indicating a strong decline in
  $\Omega_{H2}/\Omega_{HI}$ with time. The slope of the decline
  differs significantly between the compared models. Only at $z>4.0$ do the Lagos et
  al. models predict an increase in $\Omega_{H2}/\Omega_{HI}$ with
  time. The authors claim this is due to a Monte-Carlo extension of
  the merger trees to very low mass halos dominated by \hi, although
  our halo mass resolution is actually higher than theirs, so this
  seems unlikely to account for the difference with our results.  Our
  predicted evolution in $\Omega_{H2}/\Omega_{HI}$ for the
  metallicity-based \h2 recipe is much flatter compared with the
  predictions from pressure-based recipes.

\subsection{Predictions in Observation Space}
\label{sec:CO}
Our model gives predictions for the \h2 mass and surface density of
galaxies, but these are difficult to observe directly. Observations
typically use the CO luminosity as a tracer for the \h2 content of a
galaxy, assuming a CO-to-\h2 conversion factor. A proper prediction of
the CO luminosity of galaxies requires the inclusion of detailed
chemistry and radiative transfer calculations \citep{Lagos2012,Popping2013RT}. In the present work we use a CO-to-\h2 conversion
relation to convert our predicted \h2\ masses to more directly
observable CO luminosities. The advantage of working in ``Observation
Space'' is that the CO-to-\h2 conversion factor is thought to depend
on galaxy properties such as internal density and metallicity, which
are predicted by our models. Thus instead of attempting to convert CO
luminosities to \h2\ masses for the observations, we can instead make
use of our knowledge of our model galaxy properties to make a more
physically motivated galaxy-by-galaxy conversion from \h2\ to CO.

Recently, \citet{Narayanan2012} and \citet{Feldmann2012} coupled
sub-grid models of the ISM with cosmological simulations of galaxy
formation to calculate the CO-\h2 conversion factor for galaxies with
different properties. Using a coupling of an \h2-formation model and
radiative-transfer calculations to simulated isolated and starburst
galaxies, \citet{Narayanan2012} found that the average CO-\h2
conversion factor in galaxies can be represented by
\begin{equation}
\rm{X}_{\rm{CO}} = \frac{1.3\times10^{21}}{\rm{Z'}\times\Sigma_{H2}}
\end{equation}
with $X$ in units of $\rm{cm}^{-2}\,(\rm{K}\,\rm{km}\,\rm{s}^{-1})^{-1}$,
$\Sigma_{\rm{H2}}$ is the \h2 surface density in units of
$M_\odot\rm{pc}^{-2}$ and $\rm{Z}'$ is the gas metallicity in solar
units.

\citet{Feldmann2012} use a coupling of sub-grid ISM models by
\citet{Glover2011} with cosmological simulations by
\citet{Gnedin2011}. They find that, when averaged on kiloparsec
scales, the CO-\h2 conversion factor is weakly dependent on column
density and radiation field and can be described as a function of
metallicity:
\begin{equation}
\log{(\rm{X}_{\rm{CO}})} = a_1\log{(Z') }+ a_2
\end{equation}
with $a_1=-0.66$ and $a_2=20.5$ (see the Feldmann, Gnedin \& Kravtsov
2010 results averaged to 4 kpc).
\begin{figure*}
\includegraphics[width = 1.0\hsize]{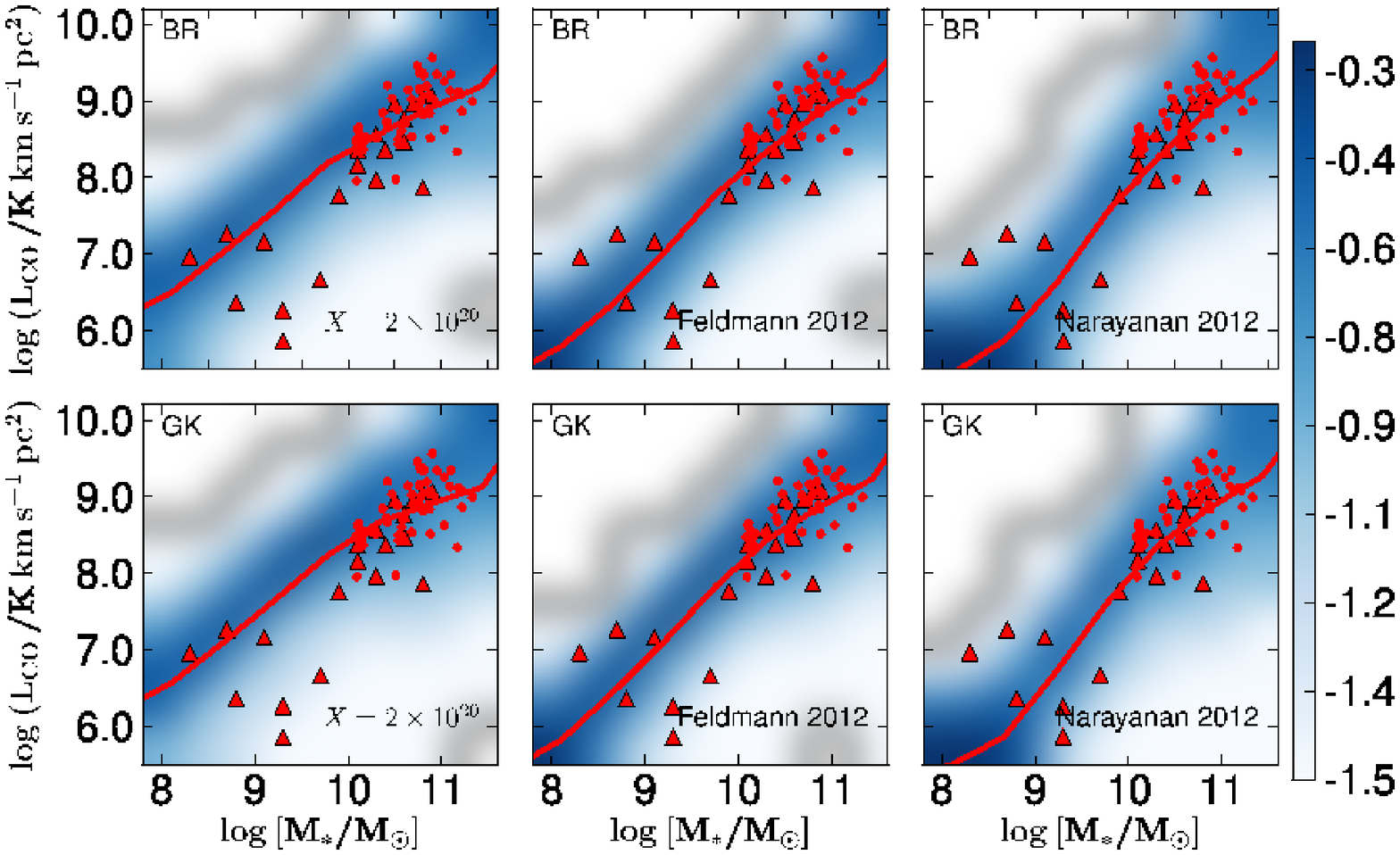}
\caption{CO(1-0) luminosity as a function of stellar mass at $z=0$
  assuming a pressure-based (top row), and metallicity-based (bottom
  row) \h2 formation recipe. We applied a fixed CO-\h2 conversion
  factor (left column), and the prescriptions suggested by
  \citet[][center column]{Feldmann2012} and \citet[][right
    column]{Narayanan2012} to calculate the conversion from \h2 masses
  to CO (1-0) luminosities. Blue shadings show the conditional
  probability distribution function $P(\rm{L}_{\rm{CO}}|M_*)$ for disc-dominated
  galaxies, whereas the red solid line shows the median fit. Red triangles and dots are literature values from \citet{Leroy2008} and
  \citet{Saintonge2011}, respectively.
\label{fig:scaling_CO}}
\end{figure*}

\begin{figure*}
\includegraphics[width = 0.9\hsize]{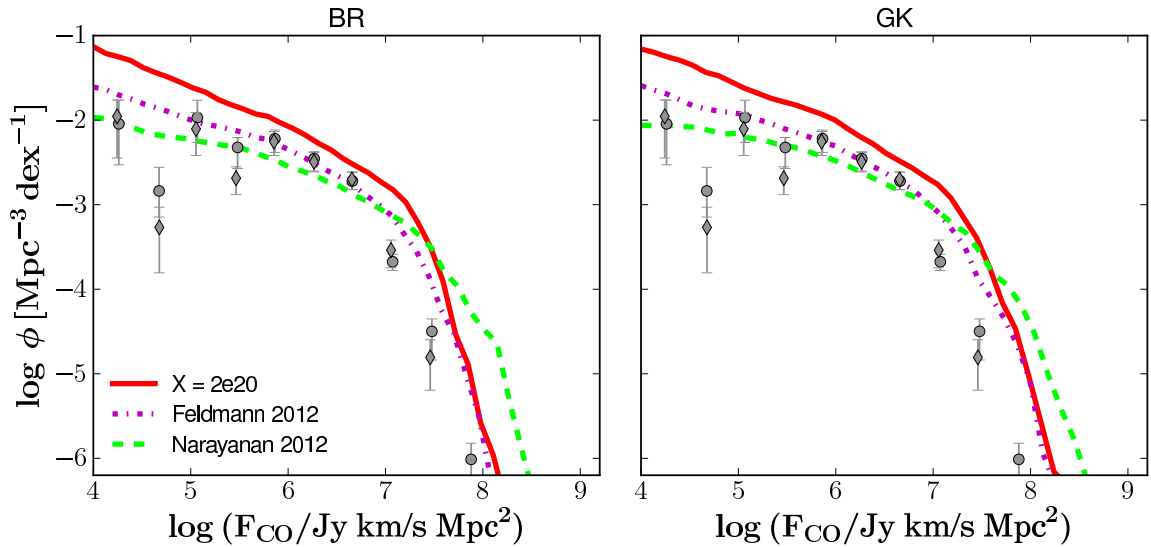}
\caption{The CO(1-0) luminosity function at $z=0$ for a pressure-
  based (left panel) and metallicity-based (right panel) \h2 formation
  model. Different line-styles represent the conversion methods from
  \h2 to CO (1-0), as detailed on the figure. The grey circles and diamonds show
  the observed CO luminosity functions from
  \citet{Keres2003}.\label{fig:cofunc_0.0}}
\end{figure*}

\begin{figure*}
\includegraphics[width = 0.8\hsize]{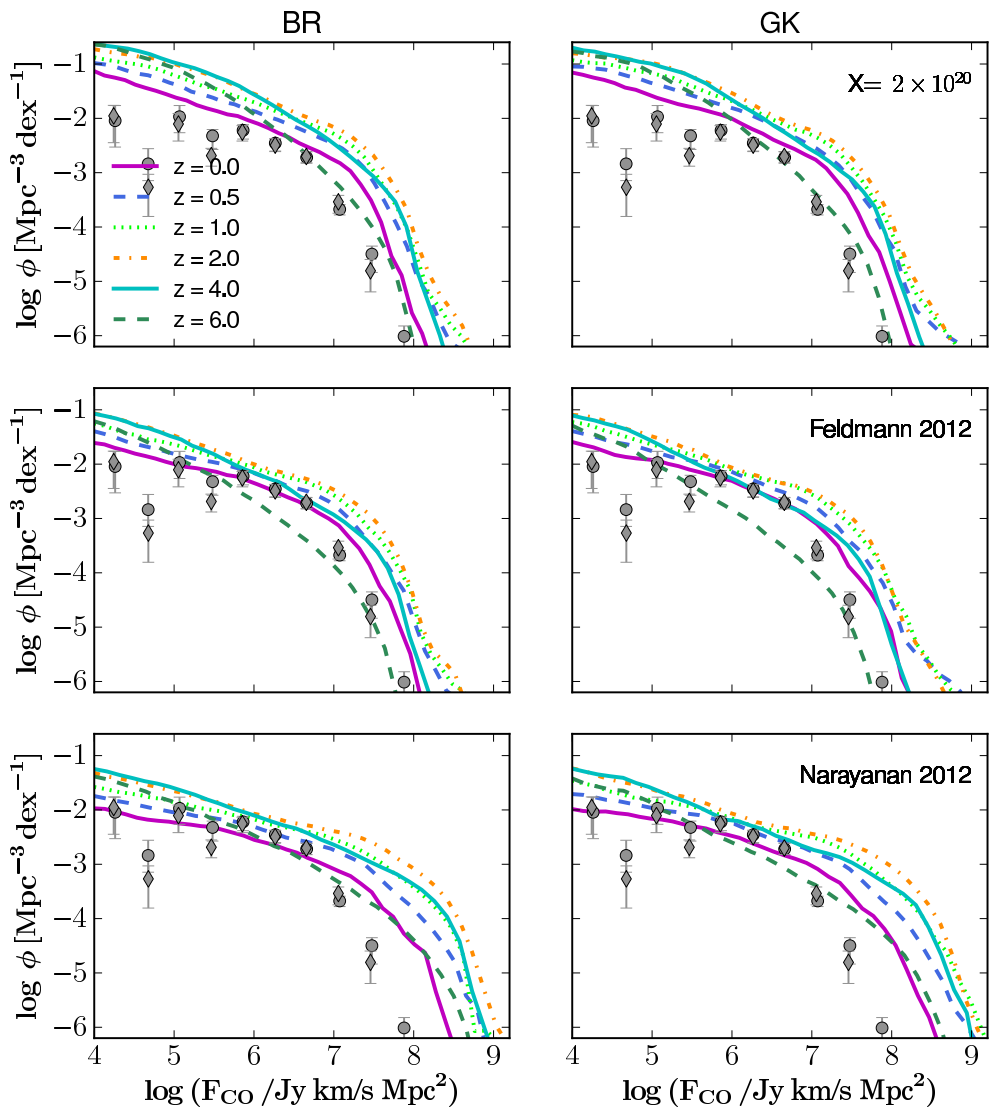}
\caption{The CO(1-0) luminosity function as a function of redshift
  assuming a fixed CO-\h2 conversion factor (top row), the
  \citet{Feldmann2012} approach (center row) and the
  \citet{Narayanan2012} approach (bottom row) for the conversion
  between \h2 mass and CO (1-0) luminosity. Left and right columns
  assume a pressure- and metallicity-based \h2 formation model,
  respectively. The grey circles and diamonds show
  the $z=0$ observed CO luminosity functions from
  \citet{Keres2003}. \label{fig:cofunc_evol}}
\end{figure*}

\begin{figure*}
\includegraphics[width = 0.85\hsize]{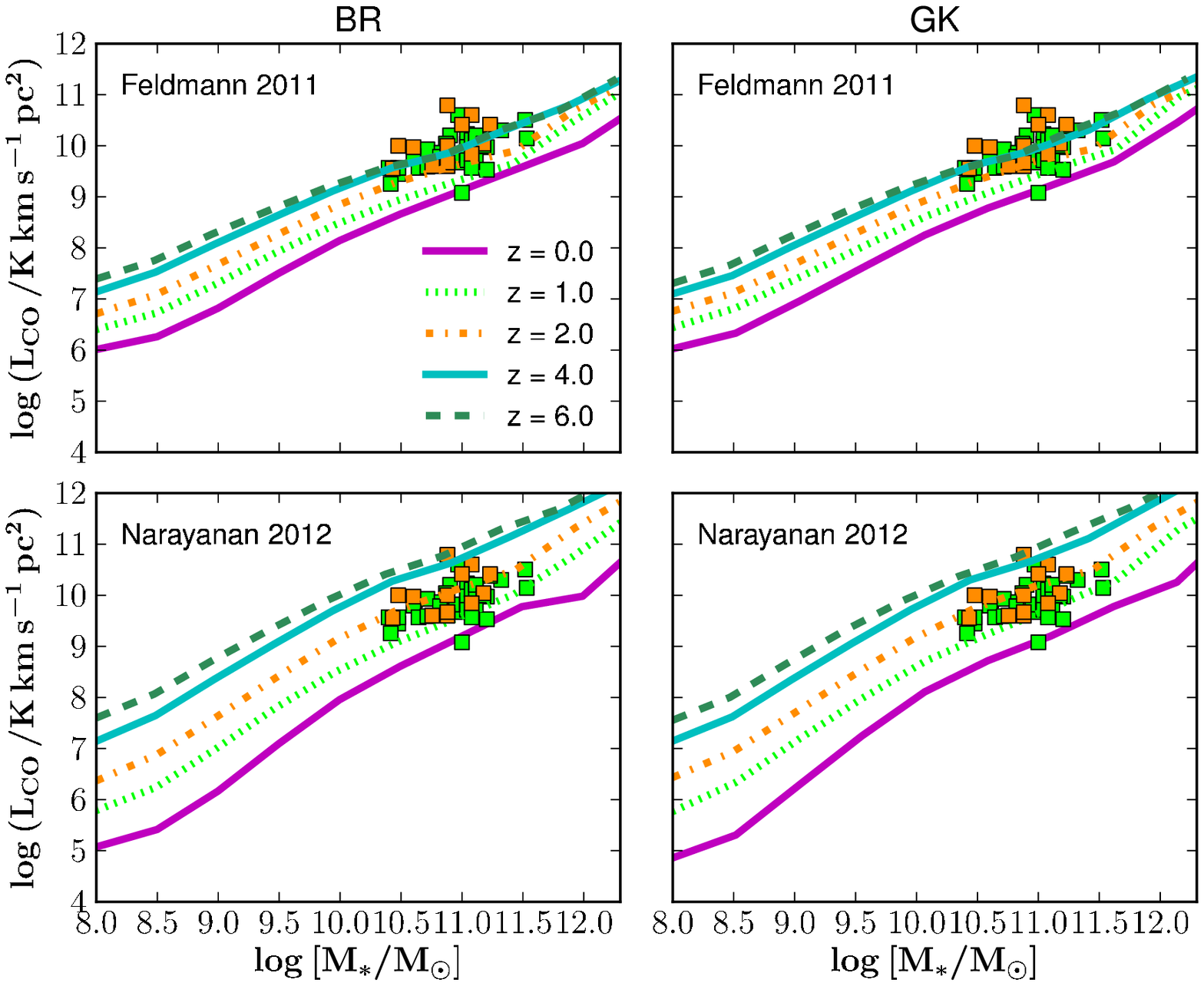}
\caption{CO (1-0) luminosity as a function of stellar mass for
  different redshift bins using a pressure- (left panel) and
  metallicity-based (right panel) star formation law. CO (1-0)
  luminosities are derived assuming the \citet[][top row]{Feldmann2012} 
  or \citet[][bottom row]{Narayanan2012}  approach for the conversion
  between \h2 and CO (1-0). Colored squares represent observations from
  \citet[and references therein]{Genzel2010} and
  \citet{Tacconi2013}.  \label{fig:mstarCO}}
\end{figure*}
We estimate the CO luminosities of our model galaxies by applying
three different assumptions for the CO-\h2 conversion factor; a fixed
conversion of $\rm{X}_{\rm{CO}} = 2 \times
10^{20}\rm{cm}^{-2}\,(\rm{K}\,\rm{km}\,\rm{s}^{-1})^{-1}$, the approach
presented by \citet{Narayanan2012} and that of
\citet{Feldmann2012}. Note that all CO luminosities presented here
correspond to the CO J$=$1-0 transition.

Figure~\ref{fig:scaling_CO} shows the CO luminosity of our model
galaxies as a function of stellar mass at $z=0.0$ for the three CO-\h2
conversion methods. Overplotted are CO luminosities observed by
\citet{Leroy2008} and \citet{Saintonge2011}. These CO luminosities
have been obtained by converting the published \h2 masses back to CO
luminosities, using the CO-\h2 conversion factor assumed in the
respective papers. The \citet{Narayanan2012} and \citet{Feldmann2012}
methods produce very similar results, and when applied to our models
both provide very good agreement with the observations. Both clearly
produce better agreement with the observations than the fixed CO-\h2
conversion factor. The slope of the relation between CO luminosity and
stellar mass varies slightly between the applied CO-\h2 conversion
method, however, this is not very well constrained by the data.

Figure \ref{fig:cofunc_0.0} shows the CO luminosity function at
$z=0.0$ obtained using the three different CO-\h2 conversion methods. The
\citet{Feldmann2012} method gives the best overall agreement with the observed CO luminosity
function.  The \citet{Narayanan2012} approach produces similar
predictions, but with a slightly shallower low-luminosity end slope
and more high CO luminosity galaxies.
A fixed conversion factor of $\rm{X}_{\rm{CO}} = 2 \times
10^{20}\,\rm{cm}^{-2}\,(\rm{K}\,\rm{km}\,\rm{s}^{-1})^{-1}$ overpredicts
the observations at all luminosities. All three methods slightly overpredict
the number of high-CO luminosity objects.

Figure \ref{fig:cofunc_evol} shows the predicted CO
luminosity-functions at redshifts 
$z=0.0$--6
for all three applied CO-\h2 conversion methods. We overplot the $z=0$
CO luminosity function obtained by \citet{Keres2003} to guide the
eye. All three CO-\h2 conversion methods yield qualitatively the
same evolutionary trends, but differ more in the details of the
predicted evolution. All models predict a relatively mild flattening
of the low-luminosity end of the CO luminosity function from $z\sim
6$--0, with a more rapid evolution on the bright end.
The Feldmann et al. and Narayanan et al. approaches give almost
identical results for the low-luminosity end, and differ more at high
luminosities. The Feldmann et al. approach predicts fewer high-CO
luminosity galaxies at high redshift.

We show the evolution with redshift of the relation between stellar
mass and CO luminosity in Figure \ref{fig:mstarCO} (assuming the
Feldmann et al. and Narayanan et al. approach for the CO-\h2
conversion factor). This diagram will, in the near future, be easily
filled with observations of the CO luminosity of galaxies from surveys
like {\it GOODS}, {\it COSMOS}, {\it CANDELS} using ALMA. As a
comparison, we plot observational results presented in
\citet{Genzel2010} and \citet{Tacconi2013}, color-coded by
redshift. We find that our model reproduces the observations very
well. There is a clear linear relation between stellar mass and CO
luminosity. The slope of this relation does not change with time and
only slightly with CO-\h2 conversion method. The normalization of the
relation does change with time, indicating that the relative amount of
CO decreases at the same rate in galaxies spanning a wide range in
stellar mass (and type). The \citet{Narayanan2012} CO-\h2 conversion
method predicts a stronger evolution in the CO luminosities with time,
driven by the dependency of the CO-\h2 conversion on the \h2 surface
density. The high surface densities in high-redshifts galaxies
decrease the CO-\h2 conversion factor, increasing the CO luminosity at
a given stellar mass. We find only minor differences between the
results obtained assuming a pressure- and metallicity-based recipe.

\section{Discussion}
\label{sec:discussion}
In this paper we have presented new predictions for the evolution of
the multiphase gas content and CO luminosity of galaxies from $z\sim
6$--0. We apply pressure- \citep{Blitz2006} and
metallicity-based \citep{Gnedin2010} \h2 formation recipes as two
different approaches to calculating the molecular fraction of cold
gas. Stars are formed following a power-law relation between the
surface density of molecular gas and the SFR surface-density
\citep{Bigiel2008}. Our goal is to assess the degree to which
observations of the gas content of galaxies at high redshift can
constrain the physics of the transformation of gas from one phase to
another, and the conversion of cold dense gas into stars. In this
section we discuss the results of this modeling effort and discuss our
findings in comparison with previous studies using similar
techniques. We will discuss the agreement and differences between the GK and BR model, draw general conclusions about the
evolution of the gas content in galaxies, provide predictions that can
help to guide future observational efforts, and discuss our results in
the context of the physics driving galaxy formation in general.

We find that both the pressure-based and metallicity-based \h2
formation recipes do well at reproducing the gas fractions and
gas-to-stellar-mass ratios of local galaxies and the trends with
stellar mass and internal galaxy density. There are only very
  small differences in the scaling relations predicted by the
  pressure- and metallicity-based recipe over the entire stellar mass
  range probed.  The predicted sizes of atomic hydrogen discs are in
good agreement with observations at $z=0$, and the sizes of the
modeled \h2 discs are in good agreement with observations in the
redshift range $z=0-2$.  We note that these recipes were taken from
empirical results calibrated to observations, or from numerical
simulations, and were not tuned to cause our semi-analytic model to
match these observations. This is an indication that, despite the
simplicity of our model for gas partitioning, SF, and disc internal
structure, we reproduce the distribution of gas in galaxies with
reasonable accuracy.

Both the pressure- and metallicity-based recipe do a fairly good job
of reproducing the \hi mass function over the whole range probed by
observations, with a small excess of high-\hi mass galaxies. Both models predict an excess of low-\hi-mass galaxies at
  $\log{(M_{\rm HI}/M_\odot)} < 8$ compared to observations.
The galaxies responsible for the excess at low-\hi masses in this model have low stellar
masses ($\log{(M_*/M_\odot)} \leq 7$) and reside in low-mass-halos
($\log{(M_{\rm{vir}}/M_\odot)} < 9-10$).  This shows that to properly
model the smallest galaxies observed in \hi, it is of key importance
to resolve halos down to masses of $\log{(M_{\rm{vir}}/M_\odot)} \sim
8$, which frequently has not been possible in previous studies. For
example, \citet{Somerville2008} presented a predicted \hi mass
function that was apparently in much better agreement with the
observed one, but this was merely an artifact of the relatively coarse
halo mass resolution ($10^{10} M_\odot$) adopted in their
simulations. Both recipes successfully predict the \h2 mass function over the entire mass range probed.

In both models, the number density of \hi-massive galaxies shows an
increase of about an order of magnitude from $z\sim 6$ to 4, then
remains nearly constant to $z\sim 0$. This result indicates that there
is a kind of self-regulated equilibrium that arises naturally in these
models. To first order, the constant high-mass end of the \hi mass function in
our models is a consequence of the balance between accretion and the
transformation of \hi into \h2. Observations have shown that \hi
saturates at surface densities of $\Sigma_{\rm{HI}} =
6-10 \, M_\odot\,\rm{pc}^{-2}$ and that higher cold gas densities are
dominated by \h2 \citep{Blitz2006,Leroy2008}. In our models, as new
gas is accreted, the amount of gas that is above surface densities
where \h2 formation is efficient increases, leading to conversion of
\hi into \h2. The constant high-mass end of the \hi mass function is a strong prediction that
can be tested by the VLA up to $z \leq 0.4$ \citep{Fernandez2013}, and
SKA and its pathfinders ASKAP and MeerKat in the near future. It will
not only probe the \h2 formation recipes, but also the physics that
drives the accretion, consumption, and heating and/or ejection of cold
gas from galaxies. 

The number density of low \h2-mass galaxies shows a strikingly
different evolution, decreasing almost monotonically from $z\sim 4$ to
$z\sim 0$. This behavior is qualitatively very similar in the two \h2
formation models. It is intriguing that this behavior --- weak
evolution of massive objects, with a decrease in the number of
low-mass objects --- is qualitatively similar to the evolution of the
observed stellar mass function \citep{Cimatti2006,Marchesini2009},
sometimes referred to as ``mass assembly downsizing''. This suggests
that ``mass assembly downsizing'' may be linked to the evolution of
the molecular gas content of galaxies and the ability to form stars
out of this molecular gas.

We also find only minor differences in the evolution of
  galaxy gas fractions between the pressure- and metallicity-based
  recipes. Gas fractions are quite high ($\gtrsim 0.7$) over a broad
range of stellar masses ($10^7 \lesssim M_* \lesssim 10^{12} M_\odot$)
from $z\sim 6$--3, then drop fairly rapidly at lower redshifts. This
drop in gas fraction occurs at higher redshift for galaxies with
higher stellar mass --- massive galaxies appear to consume or expel
their gas earlier than less massive galaxies. A similar trend holds
for the \h2 fraction of galaxies, but the rate at which the \h2
fraction drops is even faster than the rate of decline of the overall
cold gas fractions. These trends are a different manifestation of mass
assembly downsizing, and are in qualitative
agreement with the observed evolution in galaxy gas fractions
\citep{Tacconi2010,
  Magdis2012,Narayanan2012,Popping2012,Tacconi2013,Sargent2013}. Future
surveys of the molecular gas content of galaxies, as well as future
efforts to indirectly estimate the gas content of galaxies, will be
able to probe the gas content in galaxies over a much wider range in
galaxy properties and environment, improving the constraints that can
be obtained on models of galaxy formation.

In a picture where galaxy gas fractions represent the competition
between gas inflow, outflow and consumption through star formation
\citep{dave2011}, the decreasing gas fractions below redshifts of $z=3$
indicate that outflows and gas consumption largely dominate this
competition. Galaxies run out of cold gas and of molecular gas, but
not necessarily at the same rate \citep{Popping2012}. Taking into
account that galaxies form their stars out of molecular gas, this
means that declining SFRs are not only due to a decline in the cold
gas available, but also due to an even more rapid decline of the \h2
fraction of gas.

The relative \h2 content of galaxies with stellar masses below
$10^{10} M_\odot$ predicted by our models appears to be slightly too
low in the redshift regime $1.0 < z < 2.0$ compared to the predictions
by \citet{Popping2012}. This effect is still present after taking
selection criteria into account. It is probably related to the low-mass galaxy
problem in models of galaxy formation, where galaxies in this mass
regime are too passive at these redshifts with respect to the
observations. We find that the low \h2 content in these galaxies might
be driving this problem, leading to inefficient star formation. A
successful solution to the low-mass galaxy problem must also produce
higher gas fractions in low-mass galaxies at intermediate redshift.

The predictions of the cosmic-density evolution of \hi, \h2 and the
total cold gas budget show the largest differences between pressure- and
metallicity-based \h2 recipes. The metallicity-based recipe yields a
much higher cosmic density of cold gas and the density peaks at lower
redshift. More striking is the difference in the evolution of the
global \h2 fraction, $\Omega_{H2}/\Omega_{\rm gas}$. The global \h2
fraction assuming a metallicity-based \h2 formation recipe shows only
a mild decrease of a factor of $\sim 2$ from $z\sim 6$--0, whereas a
pressure-based recipe predicts a strong decrease of a factor $\sim 6$
over this redshift range. Our predicted cosmic density of \hi
$\Omega_{\rm HI}$ is, at face value, in poor agreement with estimates
from observations of DLAs for both \h2
prescriptions. However, we have presented $\Omega_{\rm HI}$ for all
galaxies without taking into account the selection criteria for
DLAs. 

It is important to note that here we have computed the global density
of gas by adding up all the gas in galaxies. However, DLAs may not
provide an unbiased estimate of the total \hi\ content of the
Universe. \citet{Berry2013} present a detailed analysis of predicted DLA
properties using the same semi-analytic models presented here, and
show that $\Omega_{\rm HI}$ derived from DLAs as in the observational
estimates shown here can differ substantially from the ``true''
underlying $\Omega_{\rm HI}$. They argue that a greater fraction of
DLAs may arise from intergalactic or circumgalactic gas at $z\gtrsim
3$, while at lower redshifts, a large amount of \hi may be in galaxies
that have column densities too low for them to be selected as DLAs,
leading to very weak evolution in $\Omega_{\rm DLA}$, as observed.

The significant differences between the metallicity- and
pressure-based recipes for \h2 formation all find their origin in low
mass galaxies ($\log{(M_*/M_\odot)} < 9$) within low mass halos
($\log{(M_{\rm{halo}}/M_\odot)} < 10$). 
A significant fraction of the cold gas and \hi that leads to the
higher cosmic densities of these quantities in the model with the
metallicity-based recipe is within virtually ``pristine'' halos that
contain less than $10^{6} M_\odot$ of stars \citep[see also the discussion in][]{Berry2013}.
These differences are driven by a lack of metals at high redshift,
necessary for the metallicity-based recipe to form molecular gas. As a
result fewer stars form, less gas is consumed and the cold gas content
of galaxies piles up. Furthermore, the lack of formed stars slows down
the production of metals necessary to form \h2. Meanwhile, the high
internal densities of high-redshift galaxies are highly conducive to
the formation of molecules through a pressure-based recipe. It is
important to note that both the pressure- and metallicity-based
recipes predict a small excess of low-\hi-mass galaxies. None
  of the various \h2 formation recipes that we have explored are able
  to remove this excess, suggesting that it may arise from other
  physical processes.

We present predictions for the CO luminosities of our modelled
galaxies using different methods to estimate the conversion between CO
and \h2. Although the general trends in CO are similar, different
approaches to estimating the CO-\h2 conversion factor yield different
predictions in detail, especially for lower-\h2-mass galaxies. The use
of a fixed conversion factor between CO and \h2 in our models
overpredicts the observed luminosity function over a wide range of CO
luminosities, although a different value for $X_{\rm CO}$ can
  change the normalization of the luminosity function. Using either the \citet{Narayanan2012} or
\citet{Feldmann2012} CO-\h2 conversion prescriptions, which depend on
galaxy properties, we obtain good agreement with the observed \h2
luminosity function below the knee, but overpredict the number of
high-\h2 mass galaxies by a significant amount, more so with the
\citet{Narayanan2012} prescription.
The predicted evolution of the CO luminosity function is qualitatively
similar to that of the \h2 luminosity function, described above,
although the detailed predictions depend somewhat on the adopted
conversion prescription. Future surveys with sub-mm and radio telescopes such as
the ALMA, PdBI, LMT, VLA, ATCA, and SKA, will be able to probe the CO $J=1-0$ luminosity function at $z \geq 2.0$
and provide valuable constraints for our models.

\subsection{Comparison with previous work}
We now discuss our results with respect to other
recent theoretical predictions of the evolution of atomic and
molecular gas in semi-analytic galaxy formation models. We will
attempt to not only point out the differences between the various
modeling efforts, but also the common results that can shed more light
on the physics at play in galaxy formation. We used the fitting
functions provided by \citet[][GK]{Gnedin2011} in our metallicity-based
recipe for the formation of \h2, whereas most previous modeling
efforts have used the analytic model of \citet{Krumholz2009}. These two
approaches have been compared and were found to be very similar except
at the lowest metallicities \citep{KrumholzGnedin}. We will
show an explicit comparison of the two approaches in SPT14, and also
find that they produce similar results. The GK fitting functions
appear to be somewhat more robust and produce better agreement with
observations, which is why we adopt them. Another difference in our
approach is that we have separated the recipes for partitioning gas
into an atomic and molecular component, and those for converting
molecular gas into stars, while in some previous works both recipes
were varied, making it more difficult to identify which aspects of the
recipes may be causing differences in the results. In
SPT14, we will present a systematic study of the
effects of varying both the gas partitioning and star formation recipes
separately. Here, we leave the star formation recipe fixed and vary
only the gas partitioning recipes.

A first attempt to study the atomic and molecular hydrogen content of
galaxies was presented in \citet{Obreschkow2009_sam} and
\citet{Obreschkow2009_letter}. The authors use the semi-analytic
predictions from \citet{DeLucia2007} and calculate the \h2 and \hi
content of galaxies in post-processing using the \citet{Blitz2006}
pressure-based formalism. This model does not include an \h2 based
star-formation recipe, but rather assumes a traditional ``total gas''
based Kennicutt star formation relation, where stars form above some
critical cold gas surface density. \citet{Obreschkow2009_sam} and
\citet{Obreschkow2009_letter} find \hi and \h2 mass functions, \h2
disc sizes and an evolution in universal density of \h2
(Fig. \ref{fig:density_ratio_evolution}) very similar to our findings
when we assume a pressure based \h2 formation recipe, and an \h2-based
star-formation recipe.

\citet{Obreschkow2009CO} estimate the CO luminosity (ranging
from CO J$=$1-0 to CO J$=$10-9) of a galaxy from its gas temperature
based on the SFR surface density or AGN bolometric luminosity under
local thermodynamic equilibrium (i.e. a single gas phase). The
authors find that the low-luminosity end of the CO J$=$1-0
luminosity function is already in place at $z=2$, contrary to our
predictions. The evolution of the bright end of the CO J$=$1-0
luminosity function is in much better agreement with our
results. \citet{Obreschkow2009CO} point out that above $z>1$ the CMB
starts to act as a bright background reducing the observed CO
J$=$1-0 luminosity. At the same time, the higher excitation
temperatures of the warm CMB in the early universe will ease the
observability of CO emission \citep{Combes1999,Gnedin2001}, although
the negative effect of the CMB dominates. These are
effects we did not include in our model but which can play a
significant role when observing young galaxies in the early
universe. In particular, sources at $z>5$ without a strong heating
by a starburst or AGN will not be detectable through low CO
transitions.

\citet{Lagos2011cosmic_evol,Lagos2011sflaw} study the evolution of the
atomic and molecular gas content of galaxies using a pressure- and
metallicity based \h2 recipe in a semi-analytic model of galaxy
formation. Their pressure-based model uses an \h2-prescription from
\citet{Blitz2006} and a star-formation model from \citet{Leroy2008},
very similar to our pressure-based model. Their metallicity-based
model follows the \h2 prescription and star-formation model presented
in \citet{Krumholz2009}. Although the authors vary the star-formation
relation in their models, the models are not calibrated to necessarily
fit the $z = 0$ luminosity functions, stellar and gas mass fractions
and mass functions. The authors find that the metallicity-based
recipes fail to reproduce the observed \hi-mass functions and select
the pressure-based recipes in combination with the \citet{Bower2006}
semi-analytic model as their preferred model. Taking into account that
the \citet{Lagos2011sflaw} models are not calibrated to match local
observations, we argue that a metallicity-based \h2 and star-formation
recipe should not be considered ruled out, although we also find (to a much lesser extent) that the metallicity-based \h2 recipe
tends to produce too many low-\hi mass galaxies at $z=0$. The KMT
model is known to break down at the lowest metallicities, due to a
failure of the assumption of chemical equilibrium in the analytic
model \citep{KrumholzGnedin}. This problem yields a rapid accumulation
of large \hi reservoirs in poor agreement with observations. We point
out that the metallicity-based recipes require the cold gas in the
initial time steps to be assigned a non-zero metallicity, otherwise no
star formation will ever take place. The results can also be somewhat
sensitive to the treatment of this ``seed'' metallicity, which may be
provided by Pop III stars.

Using their preferred model, \citet{Lagos2011cosmic_evol} find an
evolution in the \hi and \h2 mass functions, gas fractions and \h2
density of the universe very similar to our results. The authors find
a bump in the \hi mass function at $\log{(M_{HI}/\rm{h}^{-2} M_\odot)}
\sim 7.5 - 8.0$, similar to (although much larger than) the excess
number of low-\hi-mass galaxies we find. They ascribe this excess to a
mismatch between the observed and modeled radii of the galaxy
discs. We, however, have shown the sizes of the gas discs in our
models (including the sizes of the \hi and \h2 components separately)
are in good agreement with observations, so we do not think this is
the main cause of the remaining excess of intermediate \hi-mass
galaxies in our models, though it may partially explain the
  better agreement of our metallicity-based model with
  observations. \citet{Lagos2011cosmic_evol} finds a good match
between their preferred model and the observed CO luminosity function by
\citet{Keres2003}. To obtain this match the authors need to assume a
fixed CO-to-\h2 conversion factor of $X = 3.5\times10^{-20}
\rm{cm}^{-2}/\rm{K}\,\rm{km}\,\rm{s}^{-1}$.

\citet{Fu2012} also studied the redshift evolution of atomic and
molecular gas in galaxies, although the emphasis of their work lies
more on the evolution of the mass-metallicity relation. In their work
the authors use a variety of star-formation models (including the
Bigiel et al. 2008 recipe) and apply both a metallicity- \citep[based
  on][]{Krumholz2009} and a pressure-based \h2 recipe \citep[based
  on][]{Blitz2006}. Their model is calibrated to the local \hi, \h2
and stellar mass functions. The authors find that the evolution of the
atomic and molecular gas fraction of galaxies is very similar for both
applied \h2 prescriptions, and is more dependent on the star-formation
model.  Similar to our findings, their results suggest that
$\Omega_{\rm{H2}}/\Omega_{\rm{HI}}$ increases monotonically with
increasing redshift for the pressure-based \h2 recipe, whereas it
decreases at redshifts $z>3$ for the metallicity-based recipe. The
resolution of the models in \citet{Fu2012} is not sufficient to study
the differences in behavior of the low mass end of the \h2 and \hi
mass function for the pressure- and metallicity based \h2
recipes. This makes it difficult to compare our excess in low-\hi-mass
galaxies and our constraints on the different \h2 formation recipes
with the results of \citet{Fu2012}.

Despite the different implementations of the physical recipes, the
three discussed models and ours all agree that a pressure-based recipe
for \h2 formation predicts a monotonic increase for
$\Omega_{\rm{H2}}/\Omega_{\rm{HI}}$ with redshift
\citep[Fig. \ref{fig:density_ratio_evolution}, although note the
  decline in $\Omega_{\rm{H2}}/\Omega_{\rm{HI}}$ for
  the][model]{Lagos2011cosmic_evol}, whereas it flattens out for a
metallicity based recipe. We therefore conclude that the applied \h2
recipe is likely to be responsible for these trends. As discussed
extensively in the previous subsection, in the metallicity-based
models, the low metallicities at early times make \h2 formation and
hence star formation very inefficient, in spite of the higher gas
densities. Thus star formation, \h2 formation, and enrichment are
delayed in these models relative to the pressure-based models.

\section{Conclusions}
\label{sec:conclusion}
We have presented predictions for the evolution of the atomic and
molecular hydrogen content of galaxies from $z\sim 6 - 0$, based on a
semi-analytic model of galaxy formation, including new modeling of the
partitioning of cold gas in galactic discs into atomic, molecular, and
ionised phases. We present results for two different \h2 formation
recipes: one a pressure-based recipe motivated by the empirical
relation between molecular fraction and gas midplane-pressure from
\citet{Blitz2006}, and one based on numerical hydrodynamic simulations
in which the molecular fraction is highly dependent on the cold gas
metallicity as well as the local UV background \citep{Gnedin2011}. We compared our predictions to local and
high-redshift observations and adopted an alternate approach in which
we estimate the CO content of galaxies and compare directly with CO
observations. We summarize our main findings below.

\begin{itemize}
\item Without any tuning, our models correctly predict the trends
  between gas fractions and gas-to-stellar-mass ratios of \hi and \h2
  in local galaxies with mass and internal density. We furthermore
  reproduce the \hi and \h2 disc sizes of local and high
  redshift galaxies.

\item Both \h2 formation recipes reproduce the observed $z=0$ 
  \hi mass function fairly well over the whole range probed by
  observations. Both models predict a small excess of low-\hi-mass
  galaxies. The high-mass end of the \hi mass function remains remarkably
  constant at redshifts of $z  \lesssim 2.0$ for both \h2 formation recipes.

\item Both recipes correctly predict the \h2 mass function over the
  entire mass range probed. The number density of \h2-massive galaxies
  increases from $z\sim6$ to $z\sim4.0$ after which it remains fairly
  constant, whereas the number density of low-\h2 mass galaxies
  decreases almost monotonically from $z\sim4$ to $z\sim0$.

\item Galaxy gas fractions remain relatively high ($\gtrsim 0.7$) from
  $z\sim 6-3$, then drop fairly rapidly. A similar trend holds for the
  \h2 fraction of galaxies, but the drop occurs at an even higher rate.

\item The metallicity-based recipe yields a much higher cosmic density
  of cold gas over the entire redshift range probed. The cosmic \h2
  fraction as predicted by the metallicity-based recipe is much lower
  than the \h2 fraction predicted by the pressure-based recipe.

\item The galaxies responsible for the high cosmic gas density and low
  cosmic \h2 fraction all reside in low-mass
  halos ($\log{(M_{\rm{halo}}/M_\odot)} < 10$), and contain negligible
  amounts of stellar material. The build-up of atomic gas in these
  low-mass halos is driven by a lack of metals at high redshift,
  necessary to form molecular gas, stars, and produce more metals.

\item The conversion of \h2 masses to CO luminosities
  provides valuable direct predictions for future surveys with ALMA at
  low redshifts or radio interferometers such as the VLA at higher
  redshifts. None of the presented methods for the CO-to-\h2
  conversion predicts perfect agreement with observations from the
  literature, although the physically motivated nature of the
  \citet{Narayanan2012} and \citet{Feldmann2012} approaches are
  favoured over a constant CO-to-\h2 conversion factor.
\end{itemize}

The results presented in this paper can serve as predictions for future
surveys of the atomic and molecular content of galaxies. We look
forward to observations from new and upcoming facilities, that will be
able to confront our predictions, further constraining the physics
that drives the formation of molecules and the evolution of gas in
galaxies.

\section*{Acknowledgments}
We thank Michael Berry, Ann Martin, Desika Narayanan, Danail
Obreschkow, Linda Tacconi, and Martin Zwaan for providing
observational data and Marc Verheijen for stimulating discussions. We
thank the anonymous referee for many suggestions that have improved
the paper. GP
acknowledges NOVA (Nederlandse Onderzoekschool voor Astronomie) and
LKBF (Leids Kerkhoven-Bosscha Fonds) for funding and the Physics and
Astronomy department of Rutgers University for its hospitality.

\bibliographystyle{mn2e_fix}
\bibliography{references}

\end{document}